\documentclass[letterpaper,11pt]{article}
\usepackage{jheppub_mod}
\usepackage{graphicx, amsmath, amssymb, amstext}
\usepackage{relsize, float, multirow}
\usepackage{array, epstopdf, hyperref, }
\usepackage{caption, subcaption, mathtools}
\usepackage{newtxmath}
\DeclareMathOperator{\erfc}{erfc}
\DeclareMathOperator{\arcsinh}{arcsinh}

\title{ Rich structure of non-thermal
  relativistic CMB spectral distortions from high energy particle cascades
  at redshifts $z\lesssim 2\times 10^5$}
\author[a]{Sandeep Kumar Acharya,} 
\author[a]{Rishi Khatri}
\affiliation[a]{Department of Theoretical Physics, Tata Institute of 
Fundamental Research, Mumbai 400005, India}
\emailAdd{sandeepkumar@theory.tifr.res.in, khatri@theory.tifr.res.in}
\date{\today}
\abstract{ 
It is generally assumed that for energy injection before recombination, all of the injected  energy is dissipated as heat in the  baryon-photon plasma, giving rise
to the  $y$-type, $i$-type, and $\mu$-type distortions in
the CMB spectrum. We show that this assumption is incorrect when the energy
is injected in the form of energetic (i.e. energy much greater than the
background CMB temperature) particles. We evolve the electromagnetic
cascades, from the injection of high energy particles, in the
expanding Universe and follow the  non-thermal component of
\normalcolor CMB spectral distortions resulting from
the interaction of the electromagnetic shower with the background photons, 
electrons, and ions. The electromagnetic shower loses a  substantial
fraction of its energy to the CMB spectral distortions before the energy of
the particles in the shower has degraded to low enough energies that they can
thermalize with the background plasma. This spectral distortion is the
result of the interaction of non-thermal energetic electrons in the shower with the CMB and
thus has a shape that is substantially  different from the $y$-type or $i$-type distortions.  The  shape
of the final \emph{non-thermal relativistic} ($ntr$-type) CMB spectral distortion depends upon the initial energy
spectrum of the injected electrons, positrons, and photons and thus has information
about the energy injection mechanism e.g. the decay or annihilation channel
of the decaying or annihilating 
dark matter particles. The shape of the spectral distortion is also
sensitive to the redshift of energy injection. Our calculations open up a
new window into the energy injection at $z\lesssim 2\times 10^5$ which is
not degenerate with, and can be distinguished from the low redshift
thermal $y$-type distortions.}
\notoc
\begin{document}
\maketitle

\section{\label{sec:intro}Introduction}
  In the standard $\Lambda$CDM cosmology \cite{Pl2018}, the cosmic microwave background (CMB) spectrum is given by the Planck spectrum to a very good approximation. The CMB radiation decouples from the matter during the epoch of recombination at $z\sim$1100 and mostly free-streams thereafter. The ratio of the number density of photons to the number density of baryons is of the order of  $10^{9}$. A rough estimate of the redshift of recombination can be calculated by demanding that the ratio of the number density of photons energetic enough to photo-ionize neutral hydrogen atoms to the total number density of hydrogen atoms is of the order of unity. Therefore, recombination dynamics is controlled by the energetic photons in the exponential tail of the Planck spectrum while the bulk of the photons in the CMB pass through the recombination epoch unimpeded.  If there is some mechanism in the high redshift universe (before the recombination epoch) which injects energy into the tightly coupled baryon-photon fluid, it will leave its imprints in the spectral distortion of the CMB. If there is some energy injection without any additional photon creation, there can be a $y$-type or a $\mu$-type distortion depending upon the redshift at which energy was injected \citep{Sz1969,Sz19701,Is19751,Is19752,Bdd1991,Chluba:2011hw,Ks2013}. Before the recombination epoch, the background photons and baryons are tightly coupled as hydrogen and helium are ionized and their combined evolution is governed by non-relativistic Compton scattering. The scattering of photons by thermal electrons redistributes the energy of the photons modifying the photon spectrum and this process is called Comptonization. If some energy is injected at sufficiently high redshift ($z\gtrsim2.0\times 10^5$), the distorted spectrum can relax to Bose-Einstein spectrum after a sufficient number of collisions between background photons and electrons. It can not relax to a Planck spectrum since Compton scattering is a photon number conserving process. A $y$-type distortion is created at low redshifts when the Comptonization process is not efficient enough for the photon spectrum to relax to a Bose-Einstein distribution. It was emphasized in \cite{Ks2013} that, there is \emph{an extended epoch in the early universe}, corresponding to the redshift range $10^4<z<2\times 10^5$, in which an initial $y$-type distortion cannot relax to a Bose-Einstein spectrum but evolves to an intermediate-type distortion ($i$-type), the shape of which depends upon the redshift of energy injection.   \par
  \hspace{1cm} 
  There can be variety of sources which can inject energy at high redshifts like Silk damping, cosmic strings, primordial black holes, dark matter annihilation or decay etc.    \cite{Otw1986,Vv1987,Chluba:2011hw,Ksc20121,Cks2012,Sunyaev:2013aoa,Am2017}. In the tightly coupled plasma, even though the photons have a short mean free-path, they can diffuse over much larger distances through random walk. The CMB perturbations get erased on diffusion scale \citep{Silk1968} with the energy in the perturbations heating up the average CMB photons (monopole) and creating a y-type distortion \cite{Sz1970,Cks2012,Ksc20121} which can evolve to a $i$-type or $\mu$-type distortion depending upon the redshift of dissipation \cite{Ks2013}. Cosmic strings created in the phase transitions in the early universe can radiate electromagnetic radiation and particles   \cite{Otw1986,Vv1987}.
In the standard cosmological model \cite{Pl2018}, $\sim$27 percent of energy density today is attributed to the cold dark matter. Many different particle physics models with a wide range of masses for the
dark matter exist. Several direct and indirect detection searches have put stringent constraint on the allowed parameter space, though there is no conclusive detection as of yet. Though currently observable signatures of dark matter are purely gravitational, in many of the models, dark matter can annihilate or decay to short-lived standard model  particles which decay and cascade down to the lightest stable standard model particles (electron, positron, photon, neutrino, and stable hadrons).   \par\hspace{1cm}
Energy injection, around the time of recombination,
modifies the standard recombination history by injecting electrons, positrons, or
photons which can ionize and heat the background. The precise observations
of the CMB power spectrum put constraints on the energy injection history in the redshift range 500$\lesssim z\lesssim$2000 \cite{ ASS1998,Chen:2003gz,Padmanabhan:2005es,Slatyer:2009yq,Galli:2013dna,Poulin:2016anj}.
Energy injection during the dark ages (30$\lesssim z\lesssim$500) or later ($z\lesssim30$) modifies the intergalactic medium temperature, ionization and imprints its signal on global 21cm radiation \cite{Furlanetto:2006tf,Barkana:2000fd} as well as 21cm fluctuations \cite{Lz2003,Ba2004,Emf2014}, though astrophysical
uncertainty in 21cm signal prediction can be challenging. For the decaying dark
matter with lifetime short compared to the recombination epoch, there are constraints from big bang nucleosynthesis and spectral
distortions ($y$-type and $\mu$-type), \cite{SC1984,ENS1985,Hs1993,Frt2003,Poulin:2016anj,Chluba:2011hw, Ks2013}.\par
\hspace{1cm}
Almost all current calculations assume that any energy injection before the
recombination epoch goes into the heating of the background electrons as
the universe is completely ionized.  These thermal electrons then heat up
the background photons by non-relativistic Compton scattering imprinting a
characteristic spectral distortion in the CMB ($y$-type, $\mu$-type or
$i$-type) depending on the redshift of energy injection. However, there are
inevitable $y$-type distortions coming from the reionization epoch which
can mask the high-redshift $y$-distortion signal. It was realized by
Bernstein and Dodelson \cite{Bd1990} that this picture is not correct for
photons with energy $\lesssim$ keV. The basic argument is that any injected
photons of $\lesssim$ keV energies will interact with the background
electrons only through non-relativistic Compton scattering as the
photo-ionization channel is blocked when the Universe is fully
ionized. Non-relativistic Compton scattering as an energy transfer
mechanism  can be inefficient compared to the Hubble rate even at high
redshifts since in each scattering a photon loses only
$E_{\gamma}/m_{\mathrm{e}}$ fraction of its energy, where $E_{\gamma}$ is
the energy of the photon and $m_{\mathrm{e}}$ is the mass of electron. So,
these photons will survive until today and carry information about the
energy injection processes in the early Universe. They did a rough estimate
by comparing the energy transfer rate to the Hubble rate, $n_{\mathrm{e}}
\sigma_{\mathrm{T}} (E_\mathrm{thrs}/m_{\mathrm{e}})$=$H(z)$, where
$n_{\mathrm{e}}$ is the background electron density, $\sigma_{\mathrm{T}}$
is Thomson scattering cross-section, $E_\mathrm{thrs}$ is the threshold
energy of injected photon, and $H$ is the Hubble rate. Photons with energy
below $E_\mathrm{thrs}$ escape and above $E_\mathrm{thrs}$ most of the
energy goes into heating. More recently, the spectral
distortions arising from the injection of low energy photons (i.e. energy
comparable to or smaller than  the background CMB photons) were discussed
in \cite{C20151}.  \par
\hspace{1cm}
In this paper, we follow the full particle cascade starting with the high-energy injected electrons, positrons, and photons until all of the energy is dissipated as heat or escapes in the form of low energy photons or a spectral distortion of the CMB. The high energy electrons and positrons generated in the electromagnetic cascade up-scatter the CMB photons distorting the CMB spectrum. These cascading particles do not have a thermal distribution but can have relativistic energies. Therefore, we expect the distorted spectrum to be significantly different from the non-relativistic $y$-type distortion.  We will explore different specific energy injection scenarios and the ability of non-thermal relativistic ($ntr$-type) spectral distortions to distinguish between them in an upcoming paper. Since to create a spectral distortion, the high energy particle has to impart a significant energy to a CMB photon, one would expect the lightest electromagnetically interacting particles to be the main contributors. The contribution of hadrons however may not be  negligible \cite{Weniger:2013hja}. We can take the energy injection by hadrons into account by modifying the electron, positron, and photon spectrum as hadrons will give non-thermal distortion by first creating high energy electrons, positrons, or photons by interacting with the background electrons and photons. Any energy released as neutrinos would escape. We consider monochromatic electron-only, electron-positron pair, and photon-only injection. We will use as an example of application of our calculation, the decay of dark matter as a source of the initial high energy electrons, positrons, or photons. But our results are very general and are applicable for a wide variety of energy injection scenarios. In this paper, we do not consider any specific dark matter candidate. The monochromatic spectral distortion solutions can be thought of as the \emph{Green's functions} for the general energy injection problem. We can obtain the spectral distortion solution for any general initial particle spectrum by doing a linear superposition of the  monochromatic spectra calculated in this paper. \par
\hspace{1cm}
The main aim of this paper is to show that the spectral distortions from energy injection at high redshift depend upon the energy of the injected particle as well as whether the particle is an electron, a positron, or a photon. So, the spectral distortion shape is non-universal as opposed to the universal $y$-distortion spectrum. This $ntr$-type distortion can help disentangle the energy injection before recombination by new physics from the $y$-distortions created after recombination. In Sec.~\ref{sec:Motiv}, we briefly motivate our calculation and give an overview of the physics involved. Photons above $\sim$keV energy  can impart a significant fraction of their energy to the background electrons which can then produce sub-keV photons by up-scattering the CMB photons. So, the assumption that high energy photons deposit all of their energy by heating the background medium is not correct. In Sec.~\ref{sec:method}, we present our numerical approach for calculating and evolving the electromagnetic cascades starting from high energy electron, positron, and photon injection. We follow the recursive approach given in \citep{Slatyer:2009yq,Kanzaki:2008qb,Kanzaki:2009hf}. However, their calculation does not follow the evolution of low energy photons and in particular does not  take into account Doppler broadening and Doppler boosting due to the thermal motion of the background electrons and stimulated scattering. We have developed a low-energy photon evolution code to account for these processes important in determining the shape of the final spectrum. In Sec.~\ref{sec:Calc}, we show the results for spectral distortion spectrum for monochromatic electrons, photons, and electron-positron pairs with one-time injection. We show that the spectrum varies with initial injected energy upto $\sim$10 GeV for energy injection at redshift 20000. We also consider decay like energy injection profile with energy injection rate and lifetime just allowed by cosmic back-ground explorer (COBE) y-distortion limits \cite{Fixsen1996} and show that the magnitude of the $ntr$-type distortion can be of the order of $10^{-6}$ with respect to the CMB. We conclude in Sec.~\ref{sec:disc} with some comments on the future direction. We use Planck \cite{Pl2018} $\Lambda$CDM cosmological parameters with Hubble constant $H_o$=66.88 km$\mathrm{s^{-1}Mpc^{-1}}$, baryon energy density parameter $\Omega_bh^2$=0.0221, cold dark matter energy density parameter $\Omega_ch^2$=0.1206, $h=H_o$/100, Helium mass fraction ($Y_p$)=0.24.  
\section{\label{sec:Motiv}Electromagnetic cascades in the early Universe}
\begin{figure}[!tbp]
  \begin{subfigure}[b]{0.4\textwidth}
    \includegraphics[scale=1.0]{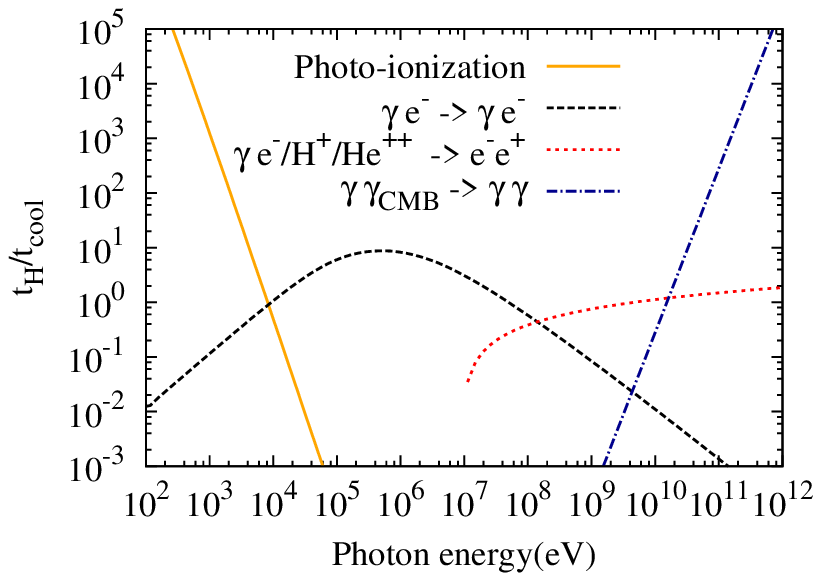}
    \caption{z=1000}
    \label{fig:hubble1z=1k1}
  \end{subfigure}\hspace{50 pt}
  \begin{subfigure}[b]{0.4\textwidth}
    \includegraphics[scale=1.0]{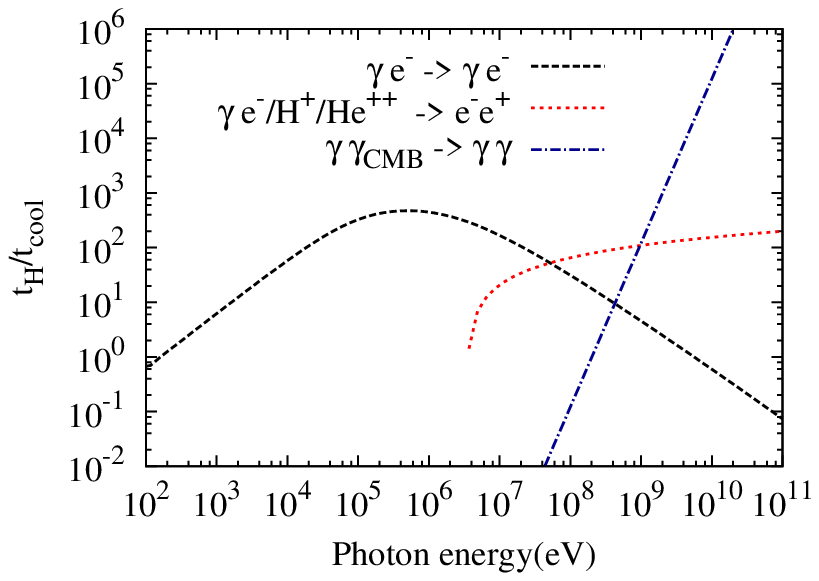}
    \caption{z=20000}
    \label{fig:hubble1z=20k1}
  \end{subfigure}
  \caption{Cooling times for the photons at different redshifts are compared with the Hubble time.}
\end{figure}
In this section, we briefly motivate the idea that is pursued in this
paper. We consider monochromatic electron, positron, and photon injection
at different redshifts and follow the electromagnetic cascade as the
injected particles interact with the background electrons, ions, and
photons. In an ionized universe, the scattering processes for
free electrons are inverse Compton scattering and Coulomb scattering. The
degradation of the injected electrons and positrons is very rapid compared
to the background expansion, so on-the-spot approximation applies
\cite{Slatyer:2009yq}. This just means that the electrons and positrons
injected at one particular redshift deposit all of their energy at that
redshift. The injection of a high energy electron or positron is equivalent
to the injection of a photon spectrum, the shape of which depends upon the
spectrum of the injected electron or positron. We will discuss the details of energy-loss mechanisms for electrons and positrons  in Sec.~\ref{subsec:enlosselec}. Since, the cooling time for photons defined by $t_{\mathrm{cool}}=1/(\mathrm{d}\ln{E_{\gamma}/\mathrm{dt}})$ \citep{Slatyer:2009yq} is comparable to the Hubble time, $t_{\mathrm{H}}$=1/$H(z)$, where $H(z)$ is the Hubble rate at redshift $z$, $E_{\gamma}$ is the energy of the photon, on-the-spot approximation is not valid for photons. Therefore, we must evolve the photon spectrum in time by following the subsequent cascade taking into account the background expansion.

  \par
\hspace{1cm}
In Fig.~\ref{fig:hubble1z=1k1}, we show the ratio of Hubble time to cooling time as a function of the energy of the injected photons at redshift 1000 with appropriate ionisation fraction \cite{Zks1969,Peebles1969,Seager:1999bc,Seager:1999km,Chluba2010,Hh2011}. In Fig.~\ref{fig:hubble1z=20k1}, we show the same at redshift 20000. For this plot there is no photo-ionization curve as the  Universe is fully ionized. For photo-ionization and pair creation on electrons, and ionized H and He, photons are destroyed to produce electrons and positrons. For photon-photon elastic scattering, the high energy photon gives a large fraction  of its energy to the background photon in one scattering. So, for these processes, the cooling time is just the inverse of collision rate, $t_{\mathrm{cool}}=\frac{1}{n_{\mathrm{t}}\sigma\mathrm{c}}$ where $n_{\mathrm{t}}$ is the number density of target particles, $\sigma$ is the  scattering cross-section and $\mathrm{c}$ is the speed of light. For Compton scattering, the cooling rate is given by the energy lost for scattering angle $\theta$, $\Delta E$, multiplied by the probability of scattering at an angle $\theta$ integrated over all angles,
\begin{equation}
 \frac{1}{t_{\mathrm{cool}}}=\frac{1}{E_{\gamma}}\int\Delta E\frac{\mathrm{d}\sigma}{\mathrm{d}\Omega}n_{\mathrm{e}}c\mathrm{d}\Omega.
\end{equation} 
  We give all the cross-sections used in Appendix \ref{app:photon}. Pair creation on CMB photons is not included in our calculations as it is not critical for determining the shape of the final CMB spectral distortion due to the following reason. Photon-photon elastic scattering and pair creation on CMB are both extremely rapid processes which degrade the initial injected photon energy \citep{Kanzaki:2008qb,Slatyer:2009yq} and boost the energy of CMB photons resulting in energetic photons. So, irrespective of the energy of the initial injected photon, we expect the spectral distortion solution to converge to a universal spectrum independent of energy, for the energies where photon-photon elastic scattering or pair creation on the CMB is the most dominant process. We will show this in Sec. \ref{sec:Calc} explicitly. \par
\hspace{1cm}
If there is some neutral hydrogen present, the low energy photons are immediately absorbed producing secondary electrons which then deposit their energy through heating, collisional excitation or ionization \cite{Ss1985}. However, if the universe is completely ionized, photoionization channel is blocked. These low energy photons can then deposit their energy into heat only via non-relativistic Compton scattering with the background electrons. As we will see, energy deposition through Compton scattering can be inefficient. \par
\hspace{1cm}
From Fig.~\ref{fig:hubble1z=20k1}, it may seem that, the energy deposition from high energy photons is quite efficient. Let's take high energy relativistic Compton scattering as an example. A high energy injected photon will give a significant amount of its energy to a background electron in a single interaction. But this high energy electron will again create lower energy photons through inverse Compton scattering on background CMB photons. If the secondary photons are of sufficiently low energy, they will deposit a small fraction of their energy by heating the background  electrons while continuously redshifting away. If they  are of high energy, they will again give a large fraction of their energy to electrons, which will again upscatter the CMB photons and the cycle repeats. So, the ultimate end result will be a distorted CMB spectrum with large amount of low energy photons created by boosting of the CMB photons from the Planck spectrum. The same story  goes for the pair-creation process. Photon-photon elastic scattering of a high energy photon on the CMB will increase the energy of the scattered CMB photon while degrading the energy of the injected photon. When the energies of these photons fall in the regime of relativistic Compton scattering or pair creation, the above argument again applies. So, the end result of the electromagnetic cascade after injection of a high energy particle will be a low energy distorted CMB spectrum which will have a different shape from the non-relativistic $y$-type or $i$-type distortion. We note that at $z\gtrsim 2.0\times 10^5$, we expect any initial spectrum to relax to a Bose-Einstein spectrum and almost all of the energy to go into $\mu$-type distortion \cite{Ks2013}. We will therefore only focus on redshifts $z\lesssim 2.0\times 10^5$. \par  \hspace{1cm}
Finally, we argue that the spectrum of low energy photons will depend on the energy of the injected particle. This is because the cross-sections of the collision processes as well as the energy exchanged between the particles at every step in the cascade are functions of the particle energies. So, we expect that the shape of the final spectral distortion will be a function of initial particle energy and retain information about the spectrum of the initial particles that are injected. 
\section{\label{sec:method}Evolution of electromagnetic cascades and low energy spectral distortions}
 To calculate the spectral distortion of the CMB, we need to evolve the electromagnetic energy cascade in an expanding universe. Electrons and photons, whether relativistic or non-relativistic, do not lose all of their energy in one collision. For example, a high energy electron can give a fraction of its energy to a CMB photon in one collision. Now we have a lower energy electron and a boosted photon with their energies fixed by collision kinematics. This lower energy electron can up-scatter another CMB photon and so on. So we have a cascade of progressively lower energy electrons and photons. To evolve the particle cascades, we follow the inductive approach worked out in Refs. \cite{Slatyer:2009yq,Kanzaki:2008qb,Kanzaki:2009hf}. \par
 \hspace{1cm}
 We divide the energy range of interest and time (or redshift) into a finite number of bins. Let's consider one particular energy bin denoted by $E_s$. The energy $E_s$ is the kinetic energy for massive particles and does not include the rest mass. This is the convention we use throughout this paper. The type of particle (electron, positron, or photon) is denoted by $\alpha,\beta$. We can write a kinetic equation for this bin as,
 \begin{equation}
 \Delta N^{\beta}_s={\sum\limits_{\alpha=e^{-},e^{+},\gamma}}\left(-\sum\limits_{j<s}P^{\beta\alpha}(E_s,E_j)N^{\beta}_s+ \sum\limits_{j>s}P^{\alpha\beta}(E_j,E_s)N^{\alpha}_j+S^{\beta}(E_s)\right),
\label{evolution} 
 \end{equation}
 where $N^{\beta}_s$ and $E_s$ are the number of particles and central energy of the bins respectively for particle $\beta$ and the bins are numbered in ascending order in energy, i.e $E_j>E_s$ for $j>s$, $\Delta N^{\beta}_s$ denotes the change in the number of particles of type $\beta$ in the corresponding energy bin in a time step, and $P^{\beta\alpha}(E_s,E_j)$ denotes the probability for a particle to transfer from higher energy bin $E_s$ to $E_j$ in that time-step. The particle  indices $\beta$ and $\alpha$ in general can be different as a particle from a high energy bin can lose its energy to a background photon or electron. These probabilities are calculated from scattering cross-sections and kinematics of various collision processes involved (see \cite{Slatyer:2009yq,Kanzaki:2008qb,Kanzaki:2009hf} for details). $S^{\beta}(E_s)$ is a source function which is non-zero if there is energy injection in the current time-step or redshift bin at energy $E_s$. So, Eq.~\ref{evolution} simply states that the change in the particle number in a particular bin is equal to the particles coming from higher energy bins minus particles lost to lower energy bins taking into account that these particles can be electrons, positrons, or photons. \par
 Eq. \ref{evolution} represents a system of coupled algebraic equations which are to be solved simultaneously. The computation simplifies due to the inductive nature of the  equations as follows. Let's consider electron-only energy injection. We choose the lowest energy bin for electron where heating by Coulomb scattering is by far the most dominant process. For electrons of energy $0<E_\mathrm{e}<<E_{CIC}$, all of the electron's energy is dissipated as heat as will be explained in Sec~\ref{subsec:enlosselec}. $E_\mathrm{e}$ is the energy of electron while $E_{CIC}$ is the cross-over energy where the energy-loss rates by Coulomb scattering and  inverse Compton scattering are equal. For an electron with energy of the order of $E_{CIC}$ and higher, we calculate the fractions of energy that go into heat by Coulomb scattering and to the CMB photons up-scattered by inverse Compton scattering. With this procedure, there is a smooth transition from Coulomb scattering to IC and we don't need to put in a hard cutoff for any process. By both these processes, the electron will lose a fraction of its energy to either the CMB or the background electrons and drop down to lower and lower energy bins until it reaches the lowest energy bin. The evolution of lower energy electrons can be used successively for higher energy electrons once they have down-scattered. This makes the calculation fast even when we have a large number of energy bins.  This method works because the energy cascade is one-way i.e from high energy to low energy and also because the probability for the high energy cascading particles to interact among themselves is negligible and they interact with only the background particles to an excellent approximation. In other words, the matrix describing Eq. \ref{evolution} is triangular and we are solving the linear algebra problem by back-substitution. Once we know the energy cascade of electrons, positrons, and photons for the monochromatic injected spectra, we can calculate the energy cascade for a more general injected spectrum by linear superposition.     
\subsection{\label{subsec:enlosselec}Energy-loss mechanisms for electrons and positrons}
For electrons, the energy loss processes are Coulomb scattering and inverse Compton scattering when the universe is fully ionized at high redshifts. We give the cross-sections and energy-loss rates used in Appendix \ref{app:elec}. 
\begin{figure}
\centering
\includegraphics[scale=1.0]{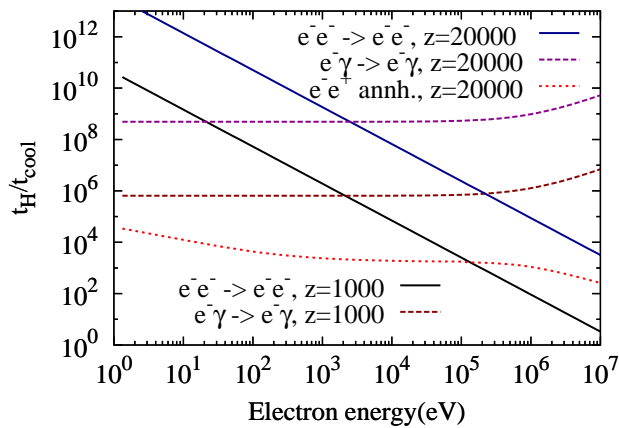}
\caption{Cooling times for electrons and positrons are compared with the Hubble time.}
\label{fig:hubble2}
\end{figure}
The cooling time for electrons compared with the Hubble time is shown in Fig.~\ref{fig:hubble2} with appropriate ionisation fraction \cite{Zks1969,Peebles1969,Seager:1999bc,Seager:1999km,Chluba2010,Hh2011} for redshifts $z$=1000 and 20000 respectively. X-axis denotes the kinetic energy of the incident electron.  For the entire energy range considered here, the cooling rate is much faster than the background expansion rate. So, an electron injected at a particular redshift deposits all of its energy as heat or to the CMB photons  immediately.  As can be seen from the plot, for electron energy $E_\mathrm{e}\lesssim$ keV, Coulomb scattering is dominant. For high energy electrons ($E_{\mathrm{e}}\gtrsim$ keV), inverse Compton (IC) scattering is dominant with cross-over at approximately 2.5 keV. The energy loss rate by IC scattering is proportional to the square of the Lorentz factor $\gamma_{\mathrm{e}}$=1.0+($E_{\mathrm{e}}$/$m_{\mathrm{e}}$). Therefore, the energy-loss rate is flat in the non-relativistic regime and starts to deviate for relativistic energies (see  Eq. \ref{IC}).  \par 
\hspace{1cm}
By IC process, the electrons boost the energy of the background CMB photons and distort the blackbody spectrum. For relativistic  electrons, the spectrum after IC scattering is calculated using results of \cite{W1979,R1995,Ek2000} with the full CMB spectrum. The formulae are given in Appendix \ref{app:elec}. The physics is the same as the non-relativistic Sunyaev-Zeldovich effect \citep{Sz1969}. However, the injected electrons can be relativistic and need not have a thermal Maxwellian distribution. Therefore, we expect the shape of the distortion to be different from the $y$-type distortion.  
\begin{figure}
\centering
\includegraphics[scale=1.0]{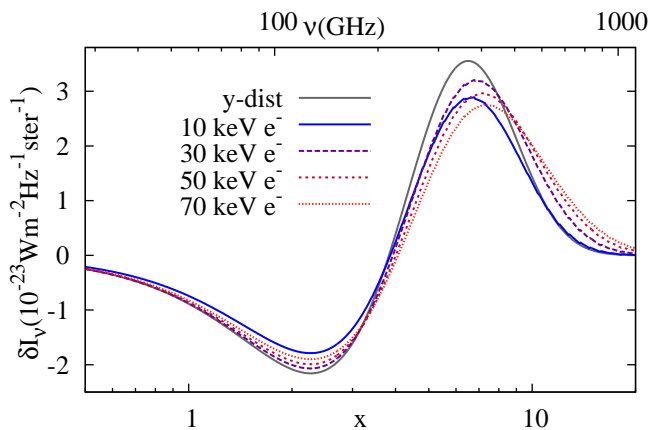}
\caption{Spectral distortion of the CMB after deposition of all of the electron energy for energy injection of $10^{-5}\times\rho_{\mathrm{CMB}}$, where $\rho_{\mathrm{CMB}}$=0.26$(1+z)^4$ eV/$\mathrm{cm^3}$ is the CMB energy density at injection redshift z=20000. We have defined dimensionless frequency $x=h\nu/kT$. The $y$-distortion curve is shown just for reference. The y-distortion from the energy dissipated as heat is not added to any of the curves.}
\label{fig:ics1}
\end{figure}
In Fig.~\ref{fig:ics1}, we plot the energy density spectrum of the
distorted CMB for various initial electron kinetic energies with
$10^{-5}\times\rho_{\mathrm{CMB}}$ of injected energy, where
$\rho_{\mathrm{CMB}}$ is the energy density of the CMB at the injection
redshift, as a function of dimensionless frequency $x=h\nu/kT$,
where $\nu$ is the frequency of the photon, $T$ is the temperature
of background electrons and photons, $k$ is the Boltzmann constant
and $h$ is the Planck constant. This is the result after the electrons have
deposited all of their energy. For any arbitrary amount of energy
injection, these plots need to be appropriately scaled while the shape
remains the same. 

As the energy of the photon and the background temperature redshifts in the
same way, a photon emitted at one particular $x$ at some high redshift will
show up at next time step with the same $x$, i.e. $x$ is invariant w.r.t the
expansion of universe. So, cosmological redshifting is implicitly taken
into account. Electrons with higher kinetic energy can impart a bigger
energy boost to the CMB photons in each scattering. Therefore, the peak of
the distorted CMB spectrum shifts to the right with the increasing electron
energy. As the collisions of high energy electrons boost the energy of the
CMB photons, moving the photons from the low energy part (Rayleigh-Jeans
region and the peak of the CMB blackbody spectrum) to the higher energy
part (Wien region) of the spectrum, the spectral intensity of the CMB in
the low energy part of the spectrum decreases while the intensity in the
higher energy part increases.  The location of the dip is determined by the
peak of the CMB blackbody spectrum since that is where most of the
background photons are, which get boosted to higher energies. As the total
injection energy is held constant, the higher energy electrons need less
number of scatterings with the CMB photons to deposit their energy. This
results in a progressive reduction of the amplitude of the dip and the peak
of the spectrum for increasing electron energy in Fig. \ref{fig:ics1}. For
the 10 keV electron, a significant fraction of the energy goes into heating
by Coulomb scattering after degradation to about ~2 keV. So, the amplitudes
of the dip and the peak for 10 keV are smaller compared to the higher
energies since we have not included the non-relativistic $y$-distortion
created by the energy dissipated as heat in the plots. The above spectrum
is the instantaneous  result after energy injection
at one redshift. This initial spectrum will get modified due to the Compton
scattering at subsequent redshifts.  We discuss the theory of evolution of
photon spectrum under Compton
scattering and  our numerical codes in the next sub-sections.  \par
\hspace{1cm}
The above result is obtained in the Thomson limit in the electron rest frame. This will not be valid for extremely high energy electrons. In the electron rest frame, the energy of the CMB photon is $E=\gamma_{\mathrm{e}} E_{\mathrm{CMB}}$ where $\gamma_{\mathrm{e}}$ is the Lorentz factor of the electron in the CMB frame and $E_{\mathrm{CMB}}$ is the energy of the CMB photon. For Thomson limit, we should have $E<<m_{\mathrm{e}}$, which will clearly be violated for high energy electrons. At higher and higher redshift, the approximation breaks down at lower and lower energies. For ultra-relativistic electrons ($\gamma_{\mathrm{e}}>>1$), Klein-Nishina cross-section must be used \cite{Gb1970}. \par
\hspace{1cm}
In Fig. \ref{fig:hubble2}, we have also shown the annihilation rate for positron with the background electrons at redshift $z$=20000. As the annihilation rate is much smaller than other scattering processes in play, we expect the positron to deposit all of its kinetic energy before annihilating with a background electron. This conclusion is also used in the works of \cite{Slatyer:2009yq,Slatyer:2015kla}. 
The energy deposition of the kinetic part of a positron's energy is identical to that of an electron when IC is the dominant energy-loss channel. When Coulomb interaction with background electrons (Bhabha scattering \cite{Bhabha1936}) becomes important, the positron is already non-relativistic with most of its energy in its mass. After losing all of its kinetic energy, it annihilates with an electron to produce two 511 keV photons. So, each injected positron is equivalent to the injection of one electron with the same kinetic energy and two 511 keV photons to a very good approximation.
\subsection{\label{subsec:enlossph}Evolution of the photon spectrum}
 We consider Compton scattering, pair creation on electrons, hydrogen and helium nuclei, and photon-photon elastic scattering with the CMB photons as the photon energy-loss processes. We have not included pair creation on the CMB photons in our calculations. This is not critical for our calculations. Photon-photon elastic scattering of high energy photons and pair creation on the CMB photons are extremely rapid processes \cite{Slatyer:2009yq} when they are important compared to the other processes. These processes will immediately degrade the energy of any photon injected within the energy range where these processes are dominant. It does not matter what the energy of the injected photon is as long as it is above the threshold of these processes. This can be seen from energy deposition curve of \cite{Slatyer:2009yq}, which converge to a single curve for energy $\gtrsim$ 1 TeV. From Fig~\ref{fig:hubble1z=20k1}, it can be seen that above 1 GeV, photon-photon elastic scattering becomes several orders of magnitude faster than the background expansion in the redshift range we are interested in. So we expect that the final spectral distortion becomes independent of the energy of the injected particle energy above a certain threshold in our case too. Indeed we will show that the final photon spectrum converges for energy of the injected particle $E\gtrsim$ 10 GeV for redshift of energy injection $z$=20000. In particular, the spectrum becomes independent of whether the injected particle is an electron, a positron or a photon. Therefore, whether the energy of the initial photon is degraded by photon-photon elastic scattering or pair-production on CMB photons becomes irrelevant and justifies our ignoring the latter process. The difference in the energy threshold in our case compared to \cite{Slatyer:2009yq} is due to the higher density and higher CMB energy at $z$=20000 compared to the recombination epoch .  \par   
\hspace{1cm}
 The non-relativistic Compton scattering can be thought of as a continuous energy loss term in the evolution equations as the energy transfer to the electron in each collision is small. So, a photon injected in one energy bin can move only to the next lower bin after losing a small amount of energy to an electron. In relativistic Compton scattering, a photon can give a sizable chunk of its energy to a background electron in a single scattering. So, it can move to any lower energy bin allowed by the Compton scattering kinematics. The electrons immediately deposit their energy via heat or to the CMB photons by IC process. Same argument applies for pair creation on matter. Photon-photon elastic scattering degrades the energy of the injected photon while boosting the energy of the CMB photon. We compare the cooling rate for all collision processes with the photon escape probability, which is a function of the Hubble rate, in each time-step and determine the fraction of photon energy going through each of the collision channels and the fraction that redshifts into the next time-step or redshift bin. We calculate the fraction of photon energy going to heat at each time-step  and subtract it from the calculation. This energy will result in non-relativistic $y$-type or $i$-type distortion depending on the redshift. We do not add this contribution to the spectral distortion plots for $ntr$-type distortions but will plot the fraction of energy going to heat separately in Sec. \ref{sec:Calc}. We obtain the photon spectrum  with redshifted photons and photons produced by IC in that time-step. We use this photon spectrum  along with the new injected photon energy as a source spectrum to be processed in next time step. The low energy photons are heating the background electrons in each time step. Our calculations show that the fractions of energy going to y-type distortion and $ntr$-type distortion are of comparable magnitude.  \par 
In non-relativistic Compton scattering, the energy-loss for photons due to recoil is proportional to the square of photon energy \cite{Il1972},
\begin{equation} 
\frac{\Delta\nu}{\nu}=\frac{-h\nu}{m_{\mathrm{e}}c^2},
\end{equation}
 where $\nu$ is the frequency of the photon. By depositing a small fraction of energy to background electrons, the photons move down in energy. The recoil formula assumes that the background electron is at zero energy. This may be a good approximation for photons with $x>>$1. But electrons do have average kinetic energy=1.5$kT$ or $x=\frac{E_{
 \mathrm{e}}}{{kT}}$=1.5. At $x\lesssim$10, the fact that the electrons have non-zero energy as well as their Maxwellian distribution  should start to play a role since the electrons in the tail of Maxwellian distribution would have energies much higher than the average. The physics can be understood as follows. Though a photon with $x\sim$10 is at higher energy compared to the average background electron energy, not every collision leads to energy-loss by the photon. Depending on the scattering angle and the energy of the electron, the photon can actually extract energy out of the background electrons. Meanwhile, the photon spectrum will broaden somewhat due to the Maxwell-Boltzmann distribution of electrons as well as due to the angle-dependent energy transfer in recoil. On average, we still expect the photons to lose their energy though it is resisted by Doppler boosting by the background electrons. We use the average energy-loss formula \cite{Pss1983}, 
\begin{equation} 
 \frac{\Delta\nu}{\nu}=\frac{4kT-h\nu}{m_{\mathrm{e}}c^2}.
\label{recoil}
\end{equation} 
 However, this formula does not capture Doppler broadening. To capture Doppler broadening, Doppler boosting and stimulated scattering accurately, we have developed a low energy code which captures the correct physics from $x\sim$0.02 to $x\sim$20 as described in the next sub-section.
\subsection{\label{subsec:lowenergyph}Evolution of the low energy photon spectrum}
We are interested in calculating accurate photon spectrum for $x\sim$0.02 to $x\sim$20 which in terms of the frequency of photons as observed today amounts to 1 GHz to 1000 GHz approximately. At frequencies $\gtrsim$ 1000 GHz, Galactic foregrounds completely obscure cosmological signatures. We would however comment on the high energy part ($x\gtrsim$20) of the spectrum in the next section.  The evolution of the photon spectrum by non-relativistic Compton scattering is given by \cite{Ss2000},
\begin{equation}
\frac{\partial p^2ndp}{\partial t} =n_{\mathrm{e}}\sigma_{\mathrm{T}}dp \int dp'\left[p'^{2}n(p')(1+n(p))P(p'\rightarrow p)\\ -p^2n(p)(1+n(p'))P(p\rightarrow p')\right],
\label{kinetic}
\end{equation}
where $n(p)$ denotes the occupation number, $p$, $p'$ denote the photon momentum,  $p^2n(p)dp$ is the number of photons in the energy bin at momentum $p$ with width $dp$, $n_{\mathrm{e}}$ is the background electron density and $\sigma_{\mathrm{T}}$ is the Thomson scattering cross-section. We are using natural units in this section with $c=\hbar=k$=1, where $c$ is the speed of light and $\hbar=\frac{h}{2\pi}$. The Kompaneets kernel $P(p\rightarrow p')$ is given by \cite{Ss2000},
\begin{multline}
P(p\rightarrow p') =\frac{1}{p}\sqrt{\frac{2m_{\mathrm{e}}}{\pi T_{\mathrm{e}}}}\left[1+\frac{p'-p}{p'+p}(1-\frac{p}{{T_{\mathrm{e}}}})\right] \left[(\frac{11}{20}+\frac{4}{5}\delta^2+\frac{2}{5}\delta^4)e^{(-\delta^2)}\right.\\ \left.  +|\delta|(-\frac{3}{2}-2\delta^2-\frac{4}{5}\delta^4)\sqrt{\frac{\pi}{4}}\erfc|\delta|\right],
\label{kernel}
\end{multline} 
where $\delta=\sqrt{\frac{m_{\mathrm{e}}}{2{T_{
\mathrm{e}}}}}\frac{p'-p}{p'+p}$, $m_{\mathrm{e}}$ is the mass of electron and ${T_{\mathrm{e}}}$ is the electron temperature. The kernel $P(p\rightarrow p')$ is the probability of a photon of energy $p$ to scatter to a photon of energy $p'$. This kernel captures the kinematics of non-relativistic Compton scattering including Doppler broadening. The factors of $(1+n(p))$ take into account stimulated scattering. Here we have assumed that the electrons see an isotropic photon distribution, so the angular information has been integrated out. The kernel $P(p'\rightarrow p)$ is related to $P(p\rightarrow p')$ by the relation,
\begin{equation}
P(p'\rightarrow p)=(\frac{p}{p'})^2 e^{\frac{p'-p}{{T_{\mathrm{e}}}}}P(p\rightarrow p').
\end{equation}
The above relation ensures that the Bose-Einstein distribution is a stationary solution to the photon evolution equation. \par
\hspace{1cm}
For this calculation we have taken 1000 log spaced bins between $x\sim$0.01 to $x\sim$35. We have made sure that we do not push the $x$ range very high  as the above kernel will not be valid anymore  at high energies where relativistic corrections become important. For our purpose, the above range is good enough. We note that at $x\sim$35, Doppler broadening and stimulated scattering will become unimportant and we can match our solution to the result of high energy calculations described in the previous sub-section which ignores these effects. As a check of our code, we inject a $y$-distortion at some redshift and follow its evolution  upto the epoch of recombination with only Compton scattering using  Eq. \ref{kinetic}. This calculation was done in \cite{Ks2013} and also in \cite{Chluba:2013vsa}. In that paper, the authors have directly solved the Kompaneets equation \cite{Kom1956} with  $y$-parameter as a proxy for time. It was shown in \cite{Ss2000} that, the kinetic equation with the kernel Eq. \ref{kernel}, under Fokker-Planck  approximation reduces to Kompaneets equation. So, essentially we are solving for the same physics under similar non-relativistic approximations. But here we directly track each individual collision instead of the combined effect of a large number of collisions described by the Kompaneets equation. For each time step, we can calculate the number of scatterings in that time step from the ratio of the Hubble rate to the collision frequency for non-relativistic Compton scattering. The number of collisions in a time-step $\Delta t$ is given by $n_{\mathrm{e}}\sigma_{\mathrm{T}}c\Delta t$ with $|\Delta t|=\frac{|\Delta z|}{(1+z)\mathrm{H(z)}}$. For the redshift range we consider, the above procedure takes a few minutes. \par
\begin{figure}
\centering
\includegraphics[scale=1.0]{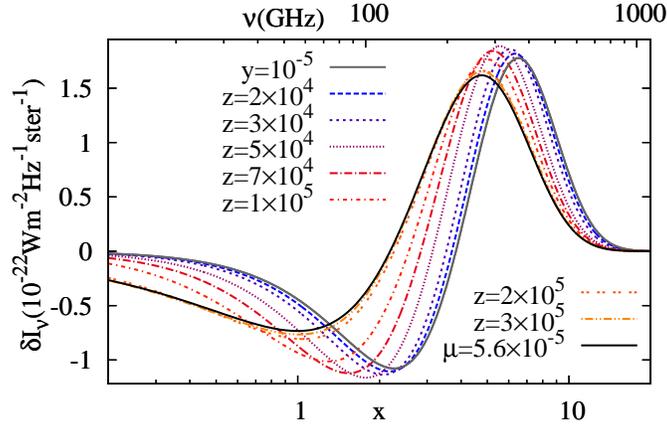}
\caption{Comptonization of y-distortion to $\mu$-distortion.}
\label{fig:ytomu1}
\end{figure}

\hspace{1cm}
We inject a $y$-distortion of magnitude $10^{-5}$ at different redshifts and follow its evolution upto hydrogen recombination redshift z=1200. The results are plotted in Fig~\ref{fig:ytomu1}. We also track the electron temperature which is given by \cite{Zl1970,Ls1971},
\begin{equation}
\frac{{T_{\mathrm{e}}}}{{T}}=\frac{\int (n+n^2)x^4dx}{4\int nx^3dx},
\label{eltemp}
\end{equation}
  where ${T}=2.725(1+z)$ K. The physics can be described as follows. Without any energy injection and ignoring the adiabatic cooling of the background electrons \cite{Zks1969,Peebles1969,Chluba:2011hw,Ksc2012}, the CMB photons and the background electrons are in thermal equilibrium. The CMB spectrum is given by the Planck spectrum which is the stationary solution of the photon evolution Eq. \ref{kinetic} with $T_{\mathrm{e}}={T}$. The magnitude of the energy injection ($10^{-5}$) seems small, but it is small only when it is compared to the CMB. The number of electrons are 9 orders of magnitude smaller compared to the number of CMB photons. So, the electrons react immediately to this energy injection with increase in temperature. The temperature increase $\frac{\Delta {T_{\mathrm{e}}}}{{T}}$ of electron immediately after the energy injection is 5.4$y$ where $y$=$\frac{1}{4}\frac{E_{\mathrm{inj}}}{E_{\mathrm{CMB}}}$ is the magnitude of the $y$-distortion. Then equilibrium between the CMB photons and the electrons is broken. The only way a new equilibrium is again reached is if the  photons redistribute their energy and attain a Bose-Einstein distribution and electrons cool to the temperature of the Bose-Einstein spectrum thus obtained. The photons cannot attain a Planck spectrum because Compton scattering is a photon number conserving process. The electron temperature at equilibrium is given by $\frac{\Delta T_{\mathrm{e}}}{{T}}=\frac{\Delta{T_{\mathrm{CMB}}}}{T_{\mathrm{CMB}}}$=2.56$y$. The final CMB Bose-Einstein spectrum has a non-zero chemical potential. We refer the reader to  \cite{Ks2013} for more details.   \par
\hspace{1cm}
\hspace{1cm}
\section{\label{sec:Calc}Calculations and results: non-thermal relativistic ($ntr$) distortions}
\begin{figure}
\centering
\includegraphics[scale=0.4]{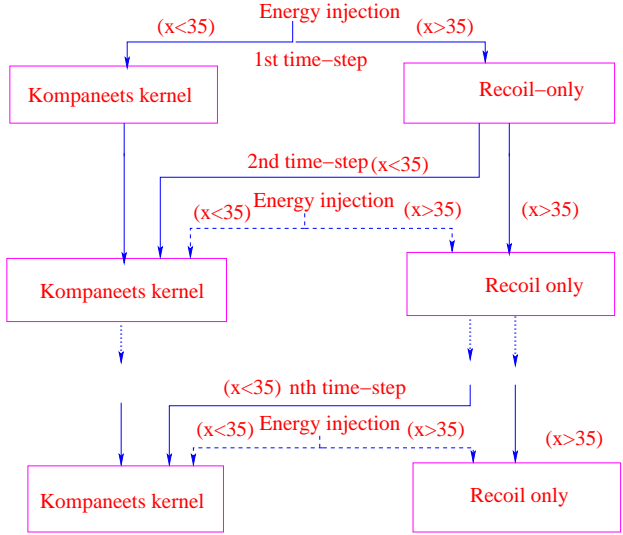}
\caption{Flowchart explaining the working and interfacing of the two codes used in the calculations}
\label{fig:flowchart}
\end{figure}
\hspace{1cm}
\begin{figure}[!tbp]
  \begin{subfigure}[b]{0.4\textwidth}
    \includegraphics[scale=1.0]{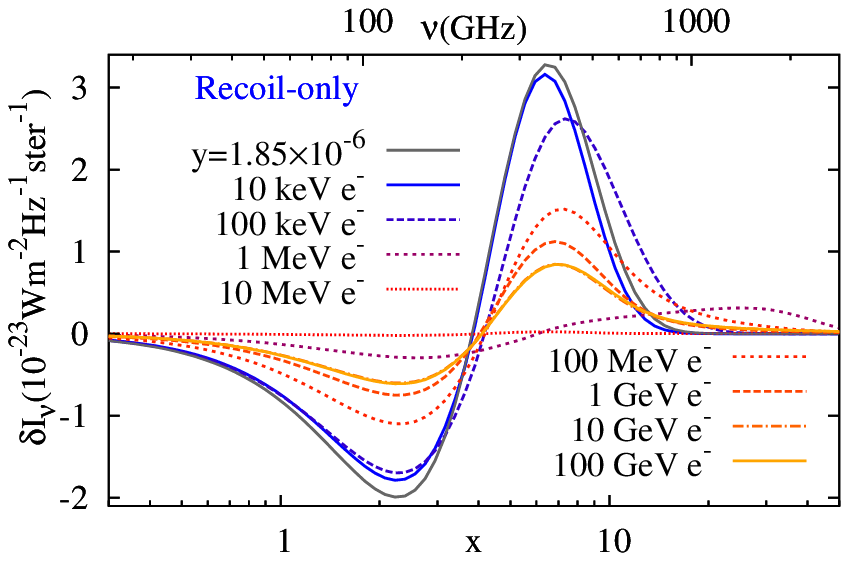}
     \caption{10 keV to 100 GeV electrons}
     \label{fig:10kev100gevrecoildelta}
  \end{subfigure}\hspace{65 pt}
  \begin{subfigure}[b]{0.4\textwidth}
    \includegraphics[scale=1.0]{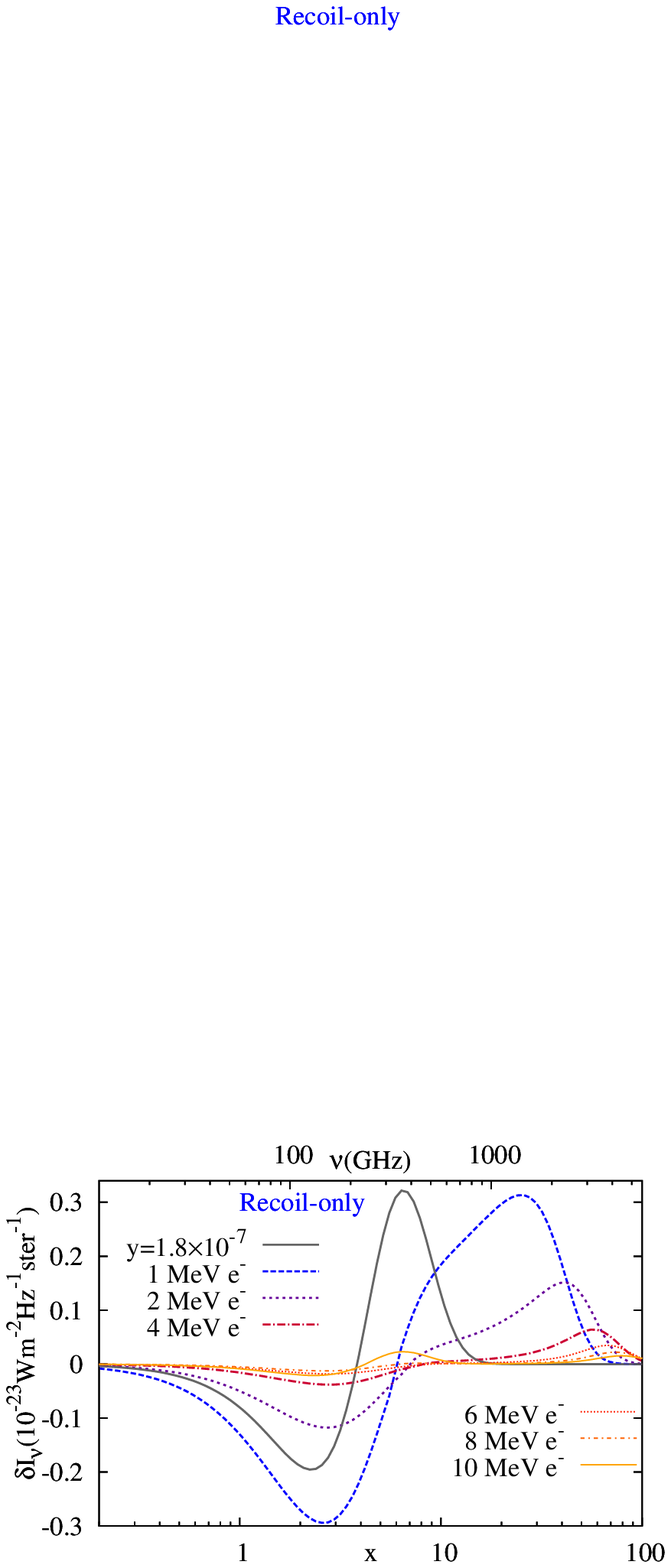}
     \caption{1 MeV to 10 MeV electrons}
    \label{fig:1mev10mevrecoildelta}
    \end{subfigure}\\
    
    \begin{subfigure}[b]{0.4\textwidth}
    \includegraphics[scale=1.0]{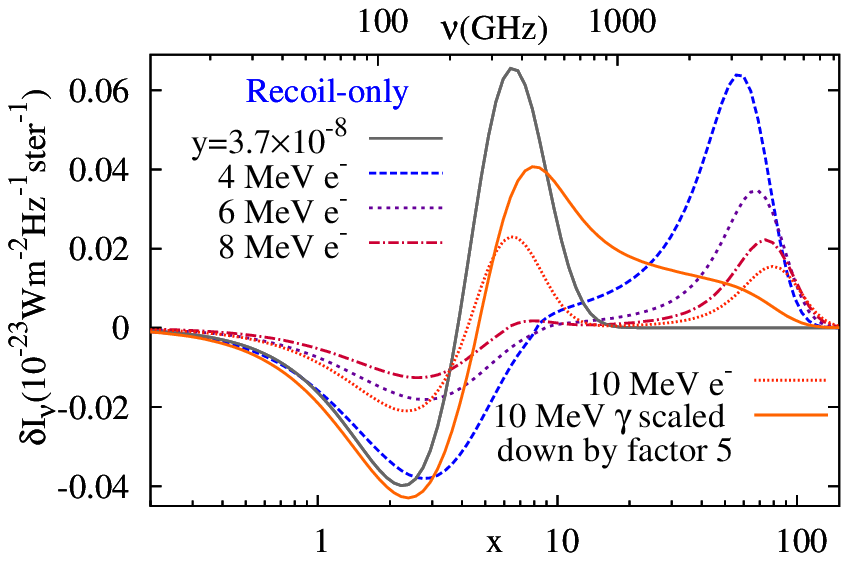}
     \caption{4 MeV to 10 MeV electrons and photons}
  \label{fig:4mev10mevrecoildelta}  
  \end{subfigure}\hspace{65 pt}
  \begin{subfigure}[b]{0.4\textwidth}
    \includegraphics[scale=1.0]{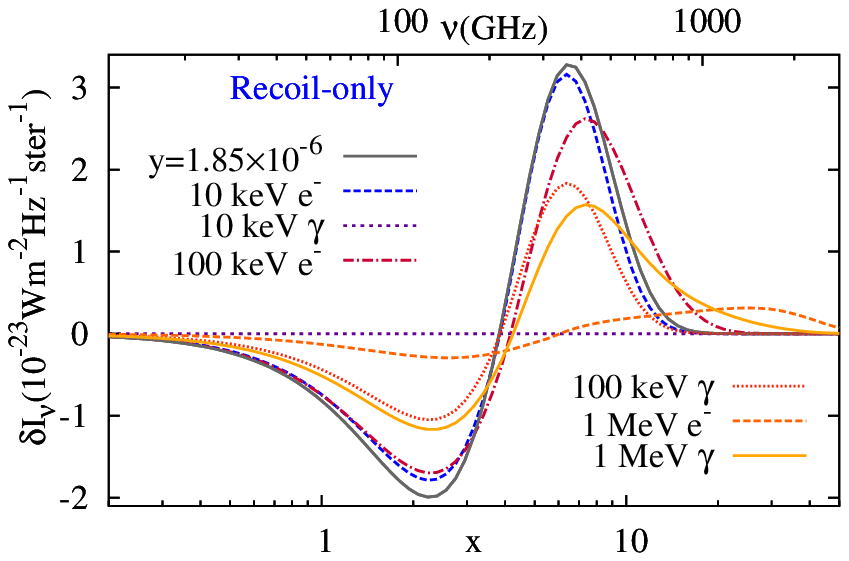}
     \caption{10 keV to 1 MeV electrons and photons}
   \label{fig:10kev10meveprecoildelta}   
   \end{subfigure}
   \caption{Comparison of spectral distortions of the CMB with varying initial electron and photon energies with \textbf{Recoil-only} code for one-time energy injection. Energy injected is $10^{-5}\times\rho_{\mathrm{CMB}}$, where $\rho_{
  \mathrm{{CMB}}}$=0.26$(1+z)^4$ eV/$\mathrm{cm^3}$ at z=20000.}
  \label{fig:10kev100gevrecoil-only}
\end{figure}
\begin{figure}[!tbp]
  \begin{subfigure}[b]{0.4\textwidth}
    \includegraphics[scale=1.0]{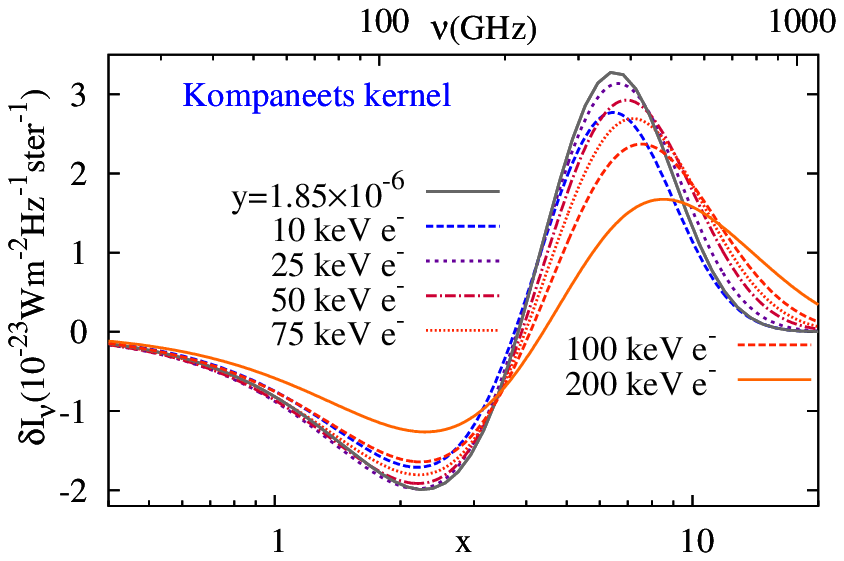}
     \caption{10 keV to 200 keV electrons}
     \label{fig:10kev200kevkomdelta1}
  \end{subfigure}\hspace{65 pt}
  \begin{subfigure}[b]{0.4\textwidth}
    \includegraphics[scale=1.0]{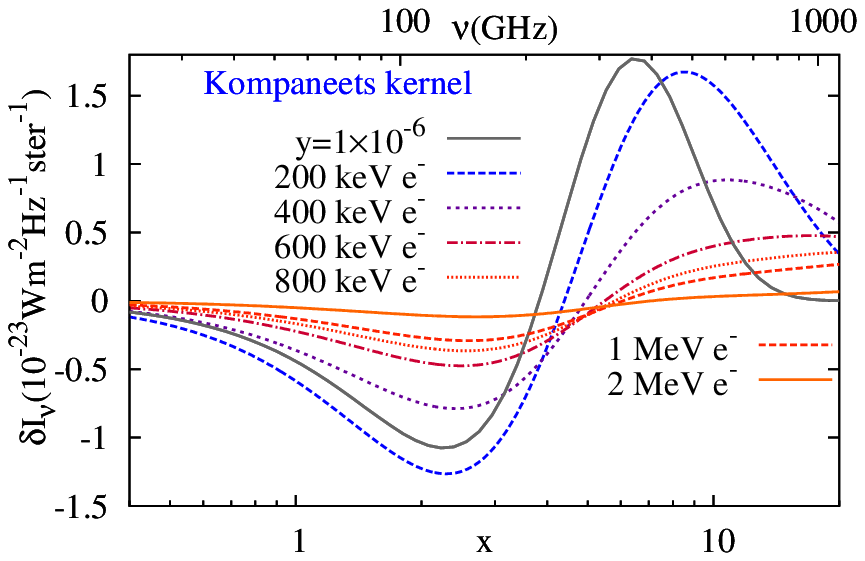}
    \caption{200 keV to 2 MeV electrons}
    \label{fig:200kev2mevkomdelta1}
    \end{subfigure}\\
    
    \begin{subfigure}[b]{0.4\textwidth}
    \includegraphics[scale=1.0]{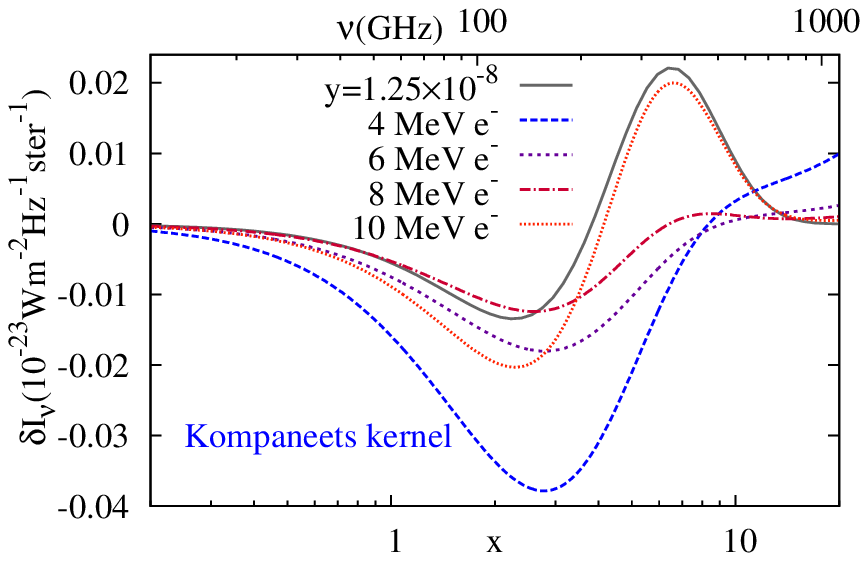}
    \caption{4 MeV to 10 MeV electrons}
   \label{fig:4mev10mevkomdelta1}  
  \end{subfigure}\hspace{65 pt}
  \begin{subfigure}[b]{0.4\textwidth}
    \includegraphics[scale=1.0]{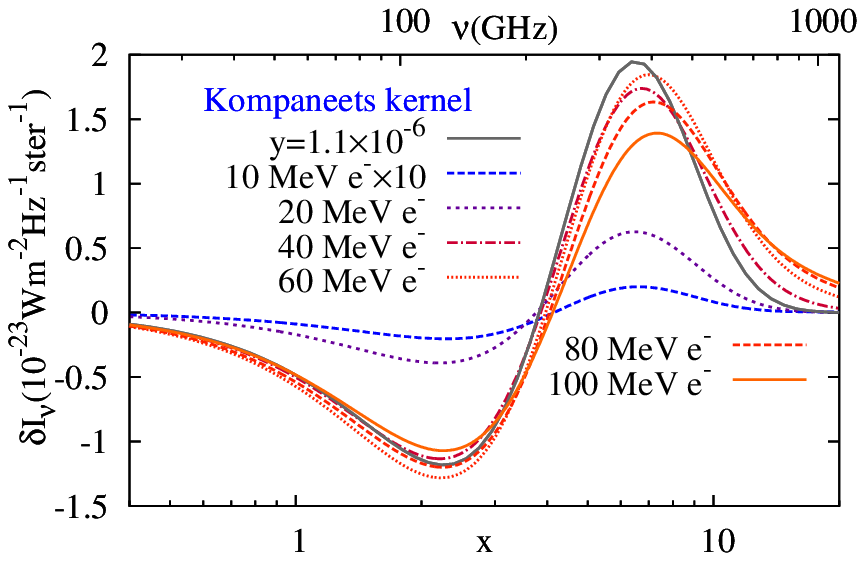}
    \caption{10 MeV to 100 MeV electrons}
   \label{fig:10mev100mevkomdelta1}   
   \end{subfigure}
   \caption{Comparison of spectral distortions of the CMB with varying initial  electron energies with \textbf{Kompaneets kernel} code for one-time energy injection. Energy injected is $10^{-5}\times\rho_{\mathrm{CMB}}$ at z=20000.}
  \label{fig:10kev100gevkom}
\end{figure}
\begin{figure}[!tbp]
  \begin{subfigure}[b]{0.4\textwidth}
    \includegraphics[scale=1.0]{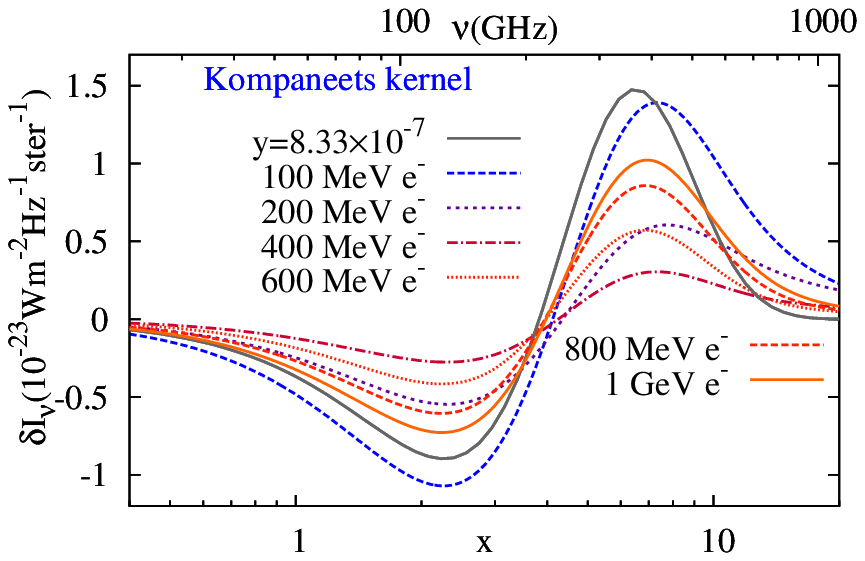}
    \caption{100 MeV to 1 GeV electrons}
     \label{fig:100mev1gevkomdelta1}
  \end{subfigure}\hspace{65 pt}
  \begin{subfigure}[b]{0.4\textwidth}
    \includegraphics[scale=1.0]{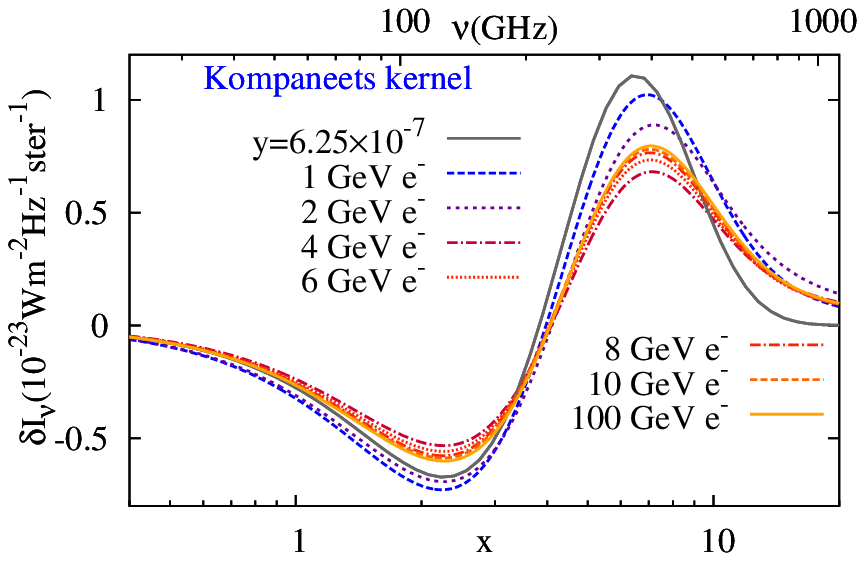}
    \caption{1 GeV to 100 GeV electrons}
    \label{fig:1gev100gevkomdelta1}
    \end{subfigure}\\
    
    \begin{subfigure}[b]{0.4\textwidth}
    \includegraphics[scale=1.0]{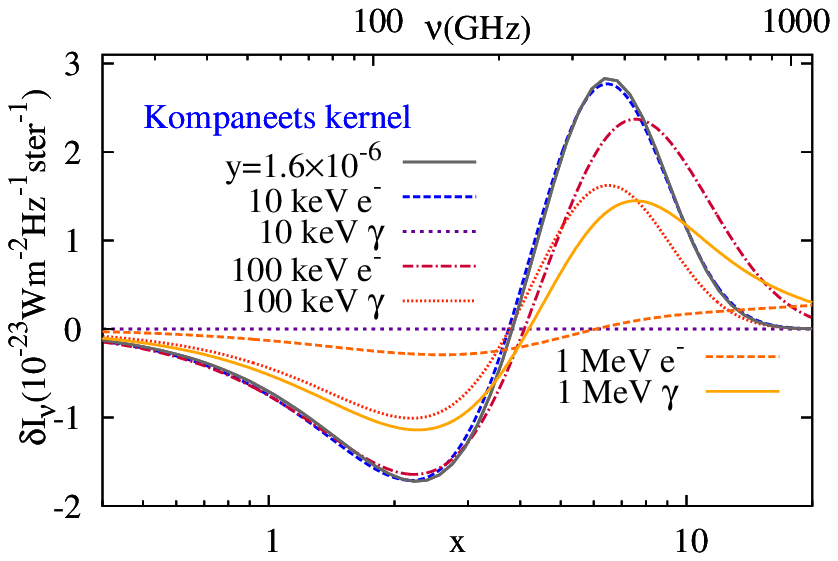}
    \caption{10 keV to 1 MeV electrons and photons}
   \label{fig:10kev10mevepkomdelta1}  
  \end{subfigure}\hspace{65 pt}
  \begin{subfigure}[b]{0.4\textwidth}
    \includegraphics[scale=1.0]{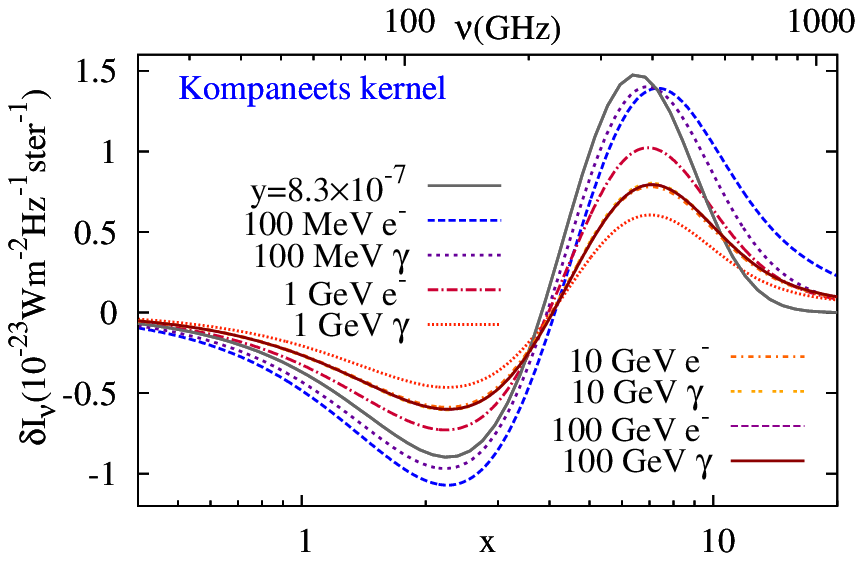}
    \caption{100 MeV to 100 GeV electrons and photons}
   \label{fig:10mev100gevepkomdelta1}   
   \end{subfigure}
   \caption{Comparison of spectral distortions of the CMB with varying initial electron and photon energies with \textbf{Kompaneets kernel} code for one-time energy injection. Energy injected is $10^{-5}\times\rho_{\mathrm{CMB}}$ at z=20000.}
  \label{fig:10kev100gevepkom}
\end{figure}
\begin{figure}[!tbp]
  \begin{subfigure}[b]{0.4\textwidth}
    \includegraphics[scale=1.0]{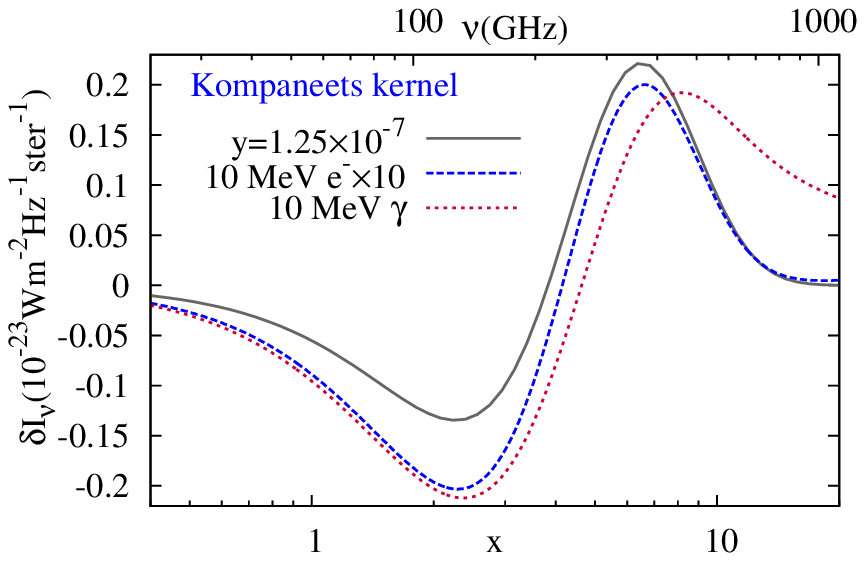}
    \caption{10 MeV electrons and photons.}
  \label{fig:10mevepkomdelta}
  \end{subfigure}\hspace{65 pt}
  \begin{subfigure}[b]{0.4\textwidth}
    \includegraphics[scale=1.0]{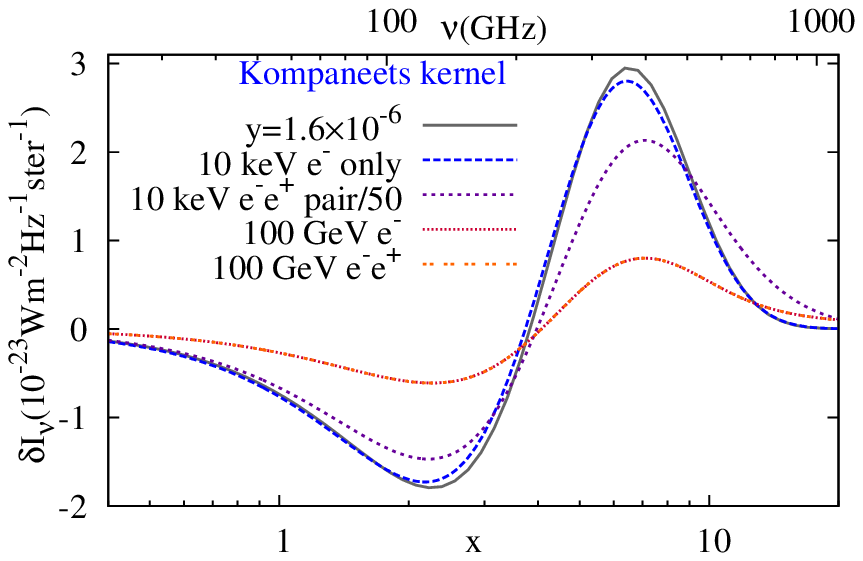}
    \caption{Electrons and electron-positron pairs.}
\label{fig:e-e+vse-delta}
    \end{subfigure}
    \caption{Comparison of spectral distortions of the CMB with \textbf{Kompaneets kernel} code for one-time injection of $10^{-5}\times\rho_{\mathrm{CMB}}$ at z=20000.}
  \label{fig:10mevepkomdelta1}
\end{figure}
\begin{figure}[!tbp]
  \begin{subfigure}[b]{0.4\textwidth}
    \includegraphics[scale=1.0]{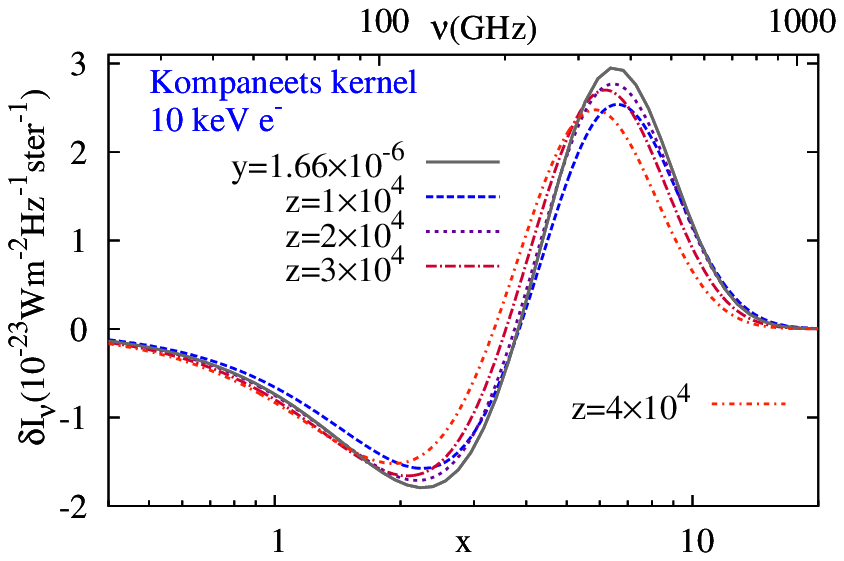}
  \end{subfigure}\hspace{65 pt}
  \begin{subfigure}[b]{0.4\textwidth}
    \includegraphics[scale=1.0]{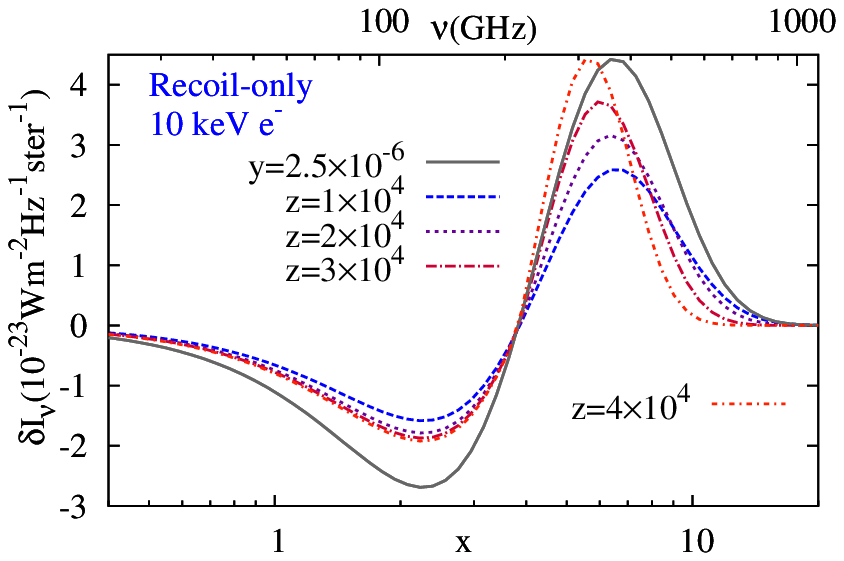}
    \end{subfigure}\\
    
    \begin{subfigure}[b]{0.4\textwidth}
    \includegraphics[scale=1.0]{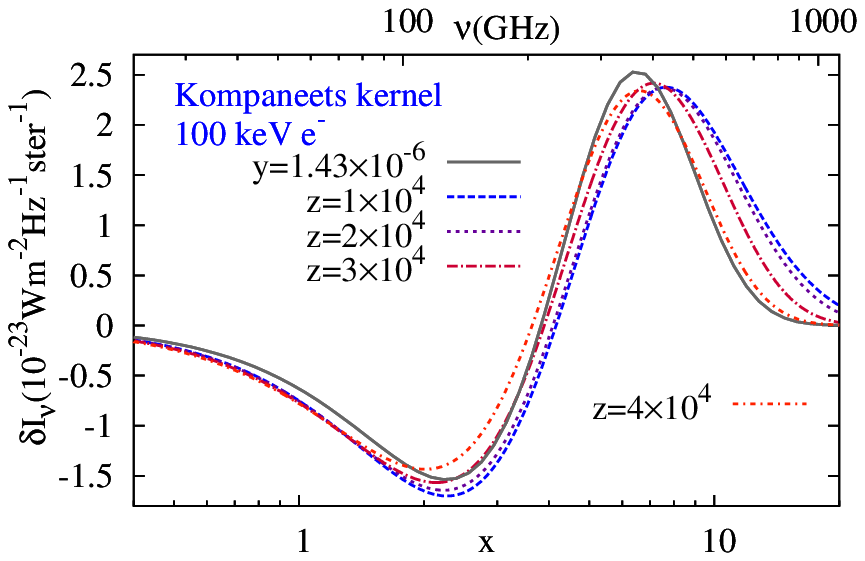}
  \end{subfigure}\hspace{65 pt}
  \begin{subfigure}[b]{0.4\textwidth}
    \includegraphics[scale=1.0]{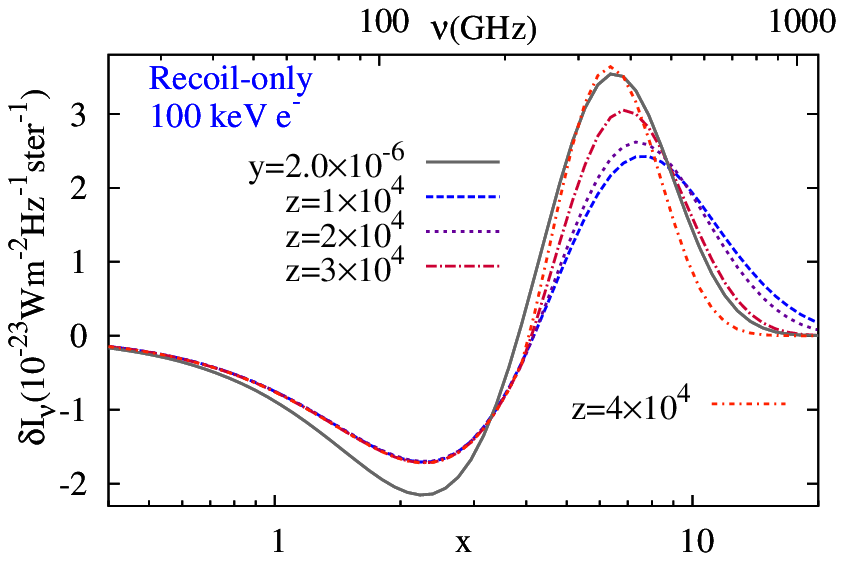}
   \end{subfigure}
   \caption{Comparison of spectral distortions of the CMB with varying redshift for one-time energy injection of $10^{-5}\times\rho_{\mathrm{CMB}}$. Top and bottom panels are for monochromatic electrons with kinetic energy 10 keV and 100 keV respectively. Left and right panels are the distortions obtained with full \textbf{Kompaneets kernel} and \textbf{Recoil-only} codes respectively.}
  \label{fig:10kev100kevvsz}
\end{figure}
\begin{figure}[!tbp]
   \begin{subfigure}[b]{0.4\textwidth}
    \includegraphics[scale=1.0]{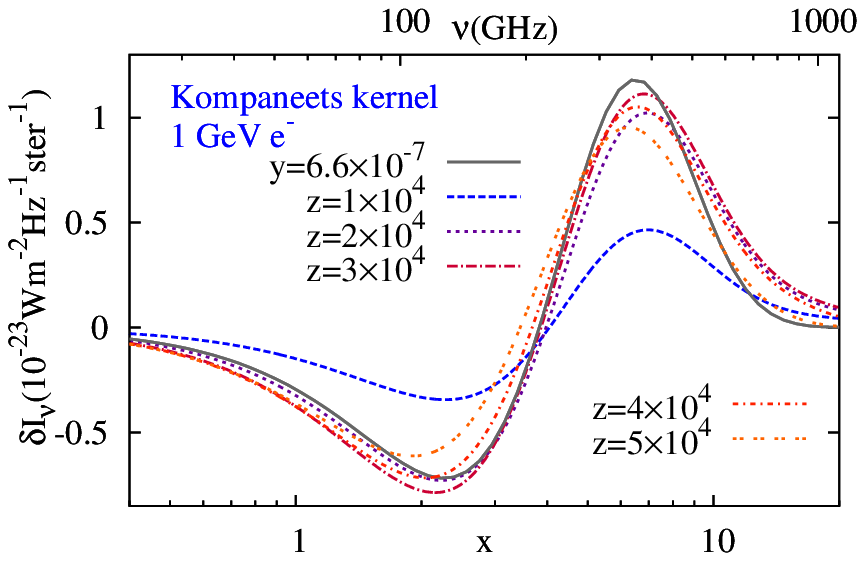}
    \label{fig:1gevvszdelta}
  \end{subfigure}\hspace{65 pt}
  \begin{subfigure}[b]{0.4\textwidth}
    \includegraphics[scale=1.0]{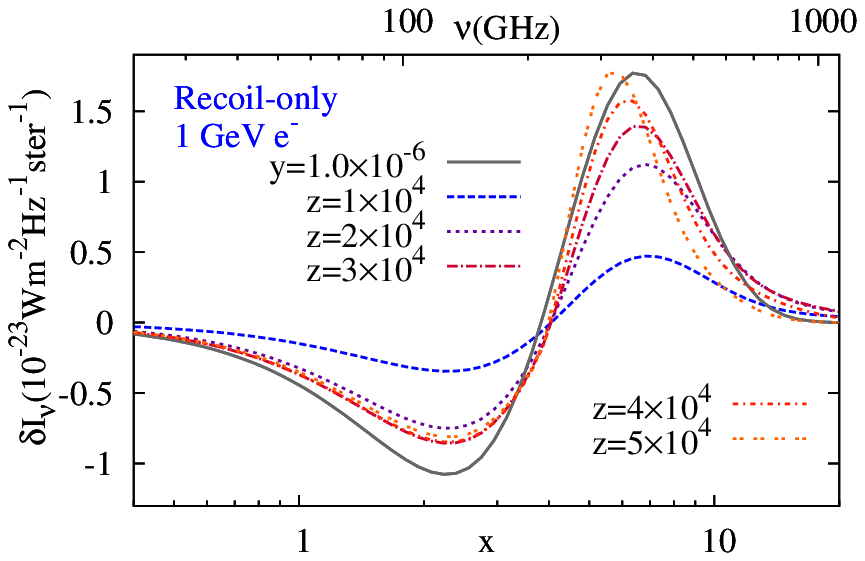}
    \label{fig:1gevvszrecoildelta}
    \end{subfigure}\\
    
    \begin{subfigure}[b]{0.4\textwidth}
    \includegraphics[scale=1.0]{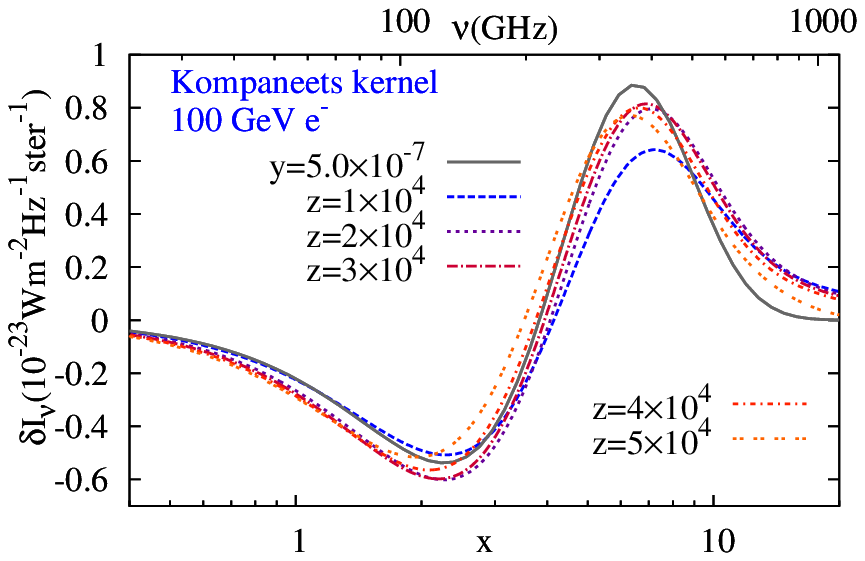}
    \label{fig:100gevvszdelta}
  \end{subfigure}\hspace{65 pt}
  \begin{subfigure}[b]{0.4\textwidth}
    \includegraphics[scale=1.0]{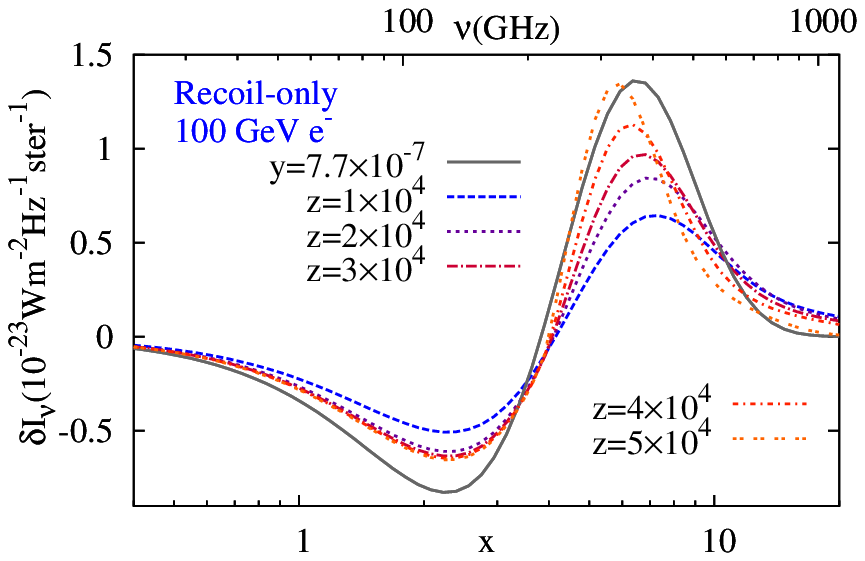}
    \label{fig:100gevvszrecoildelta}
   \end{subfigure}
  \caption{Comparison of spectral distortions of the CMB with varying redshift for one-time energy injection of $10^{-5}\times\rho_{\mathrm{CMB}}$. Top and bottom panels are for monochromatic electrons with kinetic energy 1 GeV, 100 GeV respectively. Left and right panels are spectral distortions obtained with full \textbf{Kompaneets kernel} and \textbf{Recoil-only} codes respectively.}
  \label{fig:1gev100gevvsz}
\end{figure}
\begin{figure}[!tbp]
   \begin{subfigure}[b]{0.4\textwidth}
    \includegraphics[scale=1.0]{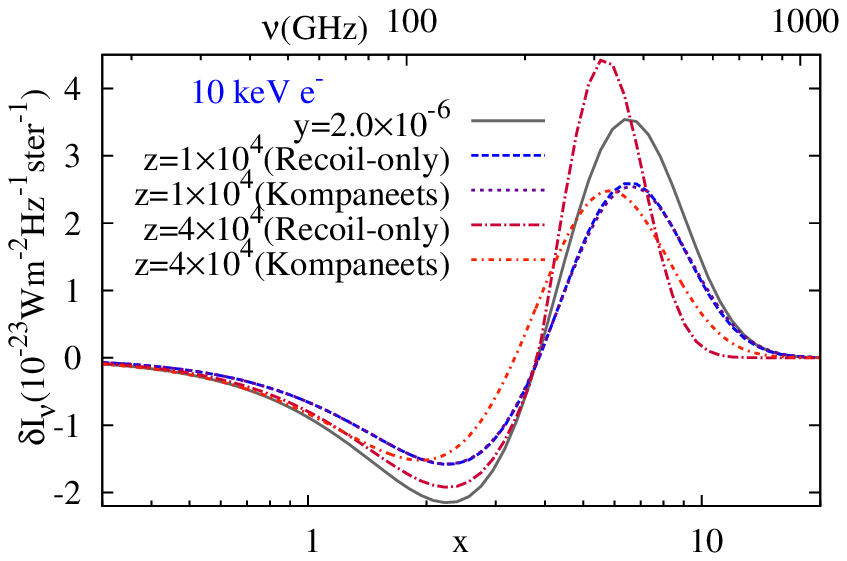}
    \label{fig:10kevvszkvsrdelta}
  \end{subfigure}\hspace{65 pt}
  \begin{subfigure}[b]{0.4\textwidth}
    \includegraphics[scale=1.0]{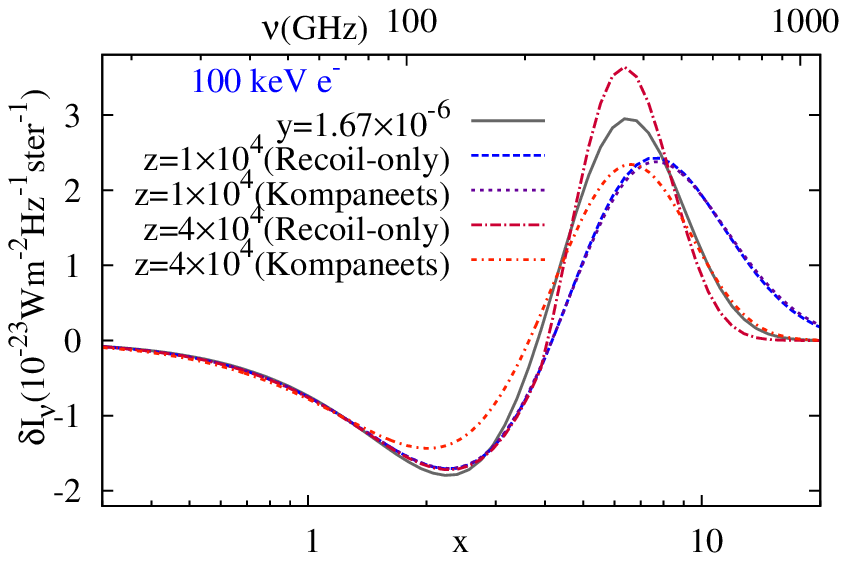}
    \label{fig:100kevvszkvsrdelta}
    \end{subfigure}\\
    
    \begin{subfigure}[b]{0.4\textwidth}
    \includegraphics[scale=1.0]{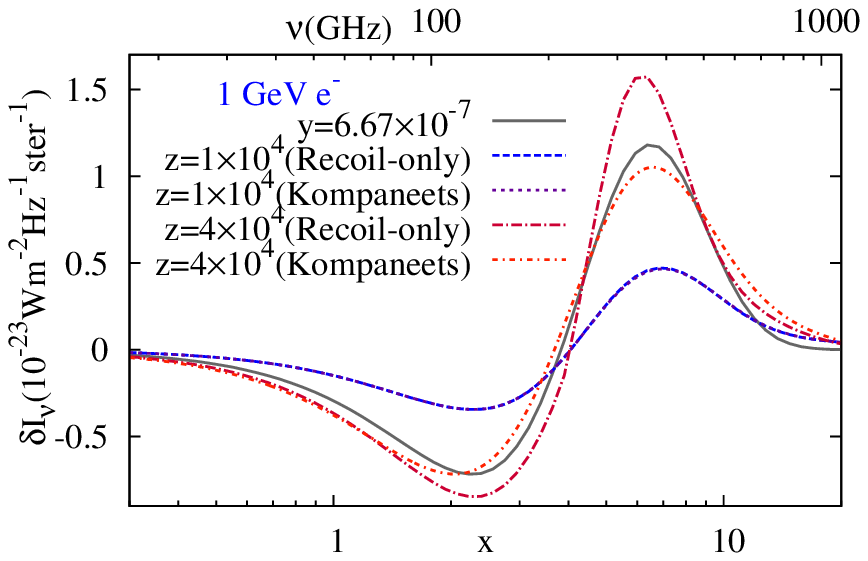}
    \label{fig:1gevvszkvsrdelta}
  \end{subfigure}\hspace{65 pt}
  \begin{subfigure}[b]{0.4\textwidth}
    \includegraphics[scale=1.0]{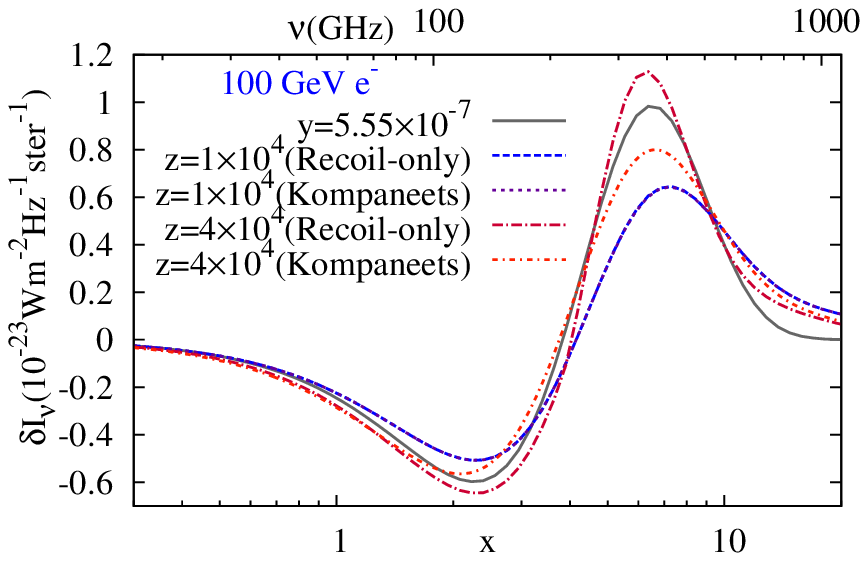}
    \label{fig:100gevvszkvsrdelta}
   \end{subfigure}
  \caption{Comparison of spectral distortions of the CMB with \textbf{Recoil-only} and full \textbf{Kompaneets kernel} solution for energy injection of $10^{-5}\times\rho_{\mathrm{CMB}}$ for one-time energy injection.}
\label{fig:10kev100gevvszkvsr}
\end{figure}
\begin{figure}[!tbp]
   \begin{subfigure}[b]{0.4\textwidth}
    \includegraphics[scale=1.0]{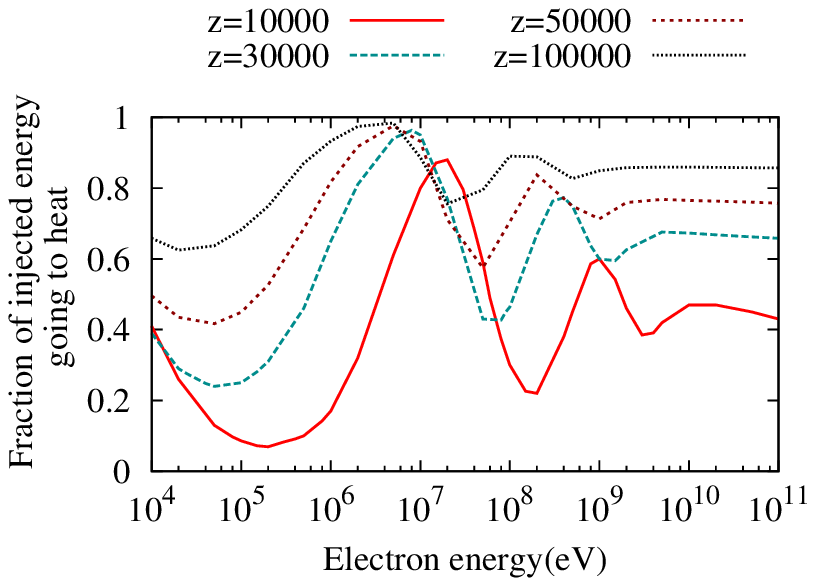}
    \caption{Fraction of energy going to heat as a function of electron energy.}
    \label{fig:elheat}
  \end{subfigure}\hspace{55 pt}
  \begin{subfigure}[b]{0.4\textwidth}
    \includegraphics[scale=1.0]{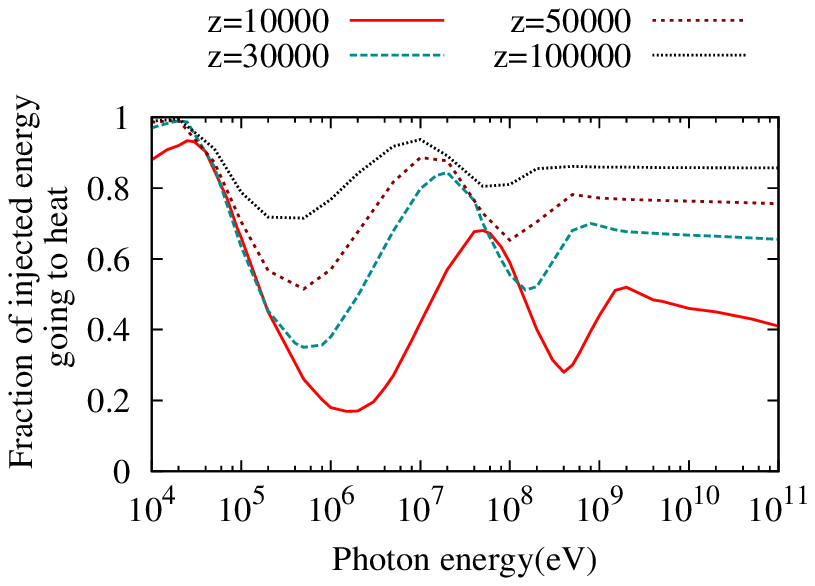}
    \caption{Fraction of energy going to heat as a function of photon energy.}
    \label{fig:plheat}
    \end{subfigure}\\
    
  \begin{subfigure}[b]{0.4\textwidth}
    \includegraphics[scale=1.0]{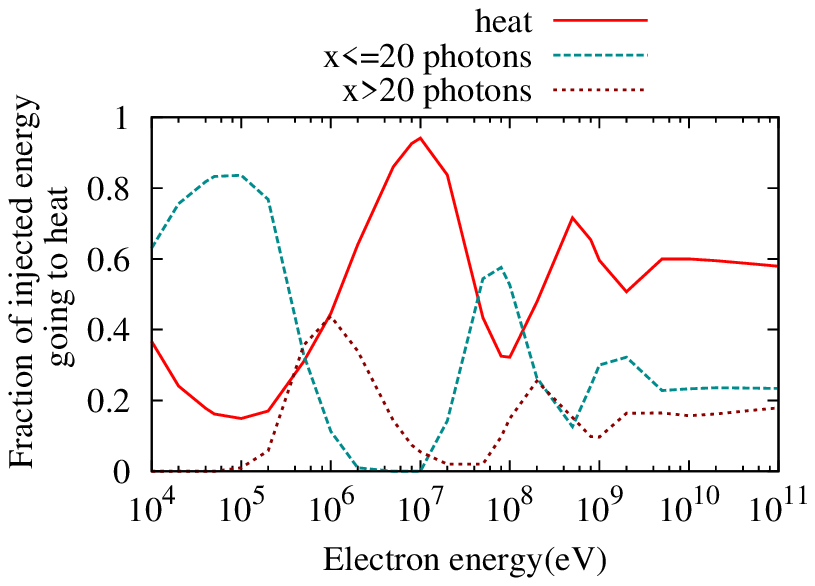}
    \caption{Fraction of energy going to heat, and photons with $x<=$20 and $x>$20 as a function of electron energy. Redshift of injection z=20000.}
    \label{fig:eheatx}
    \end{subfigure}
     \caption{Fraction of energy injected going to heat and spectral distortions.}
     \label{epheat}
     \end{figure}
     \begin{figure}[!tbp]
   \begin{subfigure}[b]{0.4\textwidth}
    \includegraphics[scale=1.0]{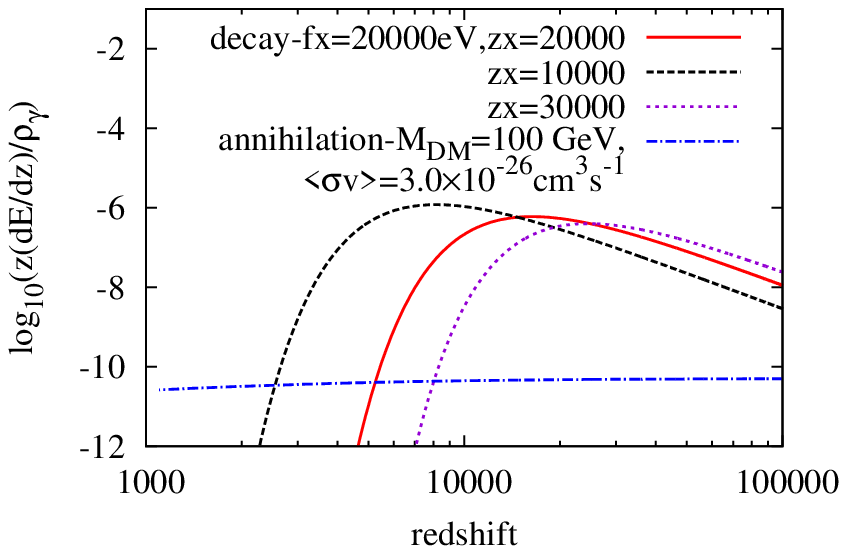}
    \caption{Ratio of energy released at a particular redshift to average CMB energy density for decay and annihilation of dark matter.}
\label{fig:decay}
  \end{subfigure}\hspace{55 pt}
  \begin{subfigure}[b]{0.4\textwidth}
    \includegraphics[scale=1.0]{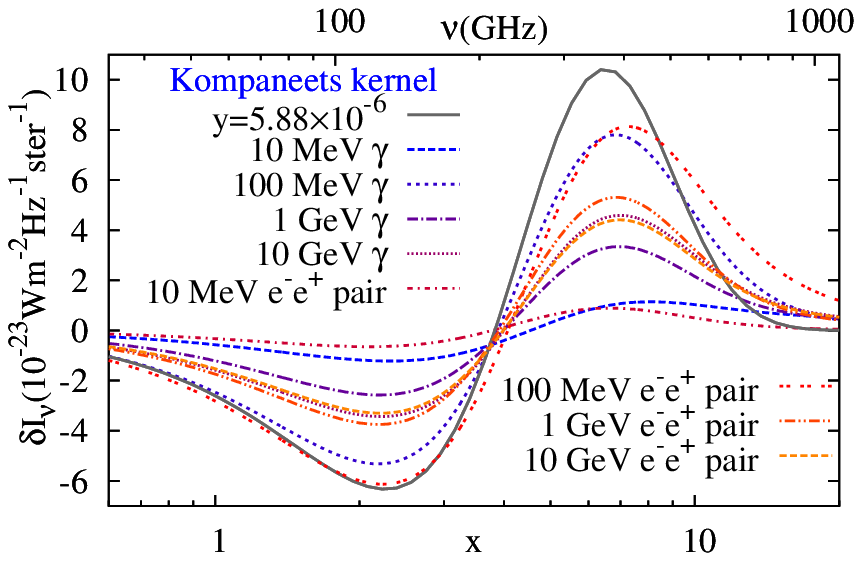}
    \caption{Spectral distortions in the CMB as a function of injected electron kinetic energy and photon energy. $z_X=2\times 10^4$, $f_X$ is chosen so that if all energy is dissipated as heat, then y=1.5$\times 10^{-5}$.}
\label{fig:10kev100gevdm}
    \end{subfigure}
    \caption{Dark matter energy injection profile and spectral distortions from dark matter decays.} 
     \label{darkmatter}
     \end{figure}
 In this section we will calculate the spectral distortions resulting from
energy injection at a single redshift.  We can linearly superpose the  resulting  spectra from energy injection at
different redshifts to get the final spectrum of any energy injection
scenario. This is because we expect the distortions to be  small
\cite{Cobe1994,Fixsen1996} and therefore the second order effect of the
energy injected at lower redshifts interacting with the distortions created
at higher redshifts to be negligibly small. 

The evolution of the photon spectrum without Doppler broadening and stimulated scattering taken into account in Compton scattering, is a good approximation for x$\gtrsim$20. We have developed a code which only takes into account the recoil effect in Compton scattering for photons of both $x\lesssim$20 and $x\gtrsim$20. These results will be denoted by \textbf{Recoil-only} in the text and figures. This calculation is similar in spirit to the calculations of \cite{Galli:2013dna,Slatyer:2015kla,Kanzaki:2008qb,Kanzaki:2009hf}. In those works, the authors were interested in deriving the constraints on the dark matter parameters by studying the effects of the energy injection on the CMB anisotropies at the recombination epoch. Since hydrogen and helium are recombining and there is significant fraction of neutral hydrogen and helium, the energy injected can be absorbed either through heating, excitation, ionization, or escape in photons with energy below 10.2eV (Lyman-alpha threshold). In our calculation, there is no excitation and ionization since we are dealing with the epoch before the  recombination era. The notion of Lyman-alpha threshold is not fundamental in our work since the  high energy photons are continually redshifting. A photon of energy 50eV emitted at z=10000 can redshift to 10eV by z$\approx$2000 (beginning of the hydrogen recombination epoch) and therefore can escape. Photon energy of 10.2eV at redshift 2000 corresponds to $x\approx$22. Even before hydrogen recombination there are two stages of helium recombination. As soon as there is a small amount of singly ionized Helium available (beginning of the recombination epoch for doubly ionized helium$\approx$8000), the part of the high energy spectrum above the Lyman-alpha threshold of He$^{+}$ will be immediately reprocessed by the recombining He$^{+}$. So, the photons with $x\gtrsim$20 will be destroyed during both hydrogen and helium recombination which can itself lead to interesting spectral distortions \cite{Chluba:2008aw}. The important thing is that photons and spectral distortions at $x\lesssim$20 survive through recombination almost unaffected and can therefore be observed today. \par
\hspace{1cm}
For photons with $x\lesssim$20, we take into account Doppler broadening and
stimulated scattering with our low energy evolution code. These results
will be denoted by \textbf{Kompaneets kernel} in the text and figures. The
low energy photons can directly come from energy injection or from
production of secondary  photons from the high energy electrons and photons
which are tracked by our  \textbf{Recoil-only} code. So, the
\textbf{Kompaneets kernel} code has to be interfaced with
\textbf{Recoil-only} code as described below. At the start of  a time-step,
the $x>$35 portion of the injected spectrum from the previous timestep is
evolved with \textbf{Recoil-only} code to get the spectrum for $x<$35 at
the end of current time-step to be processed by the \textbf{Kompaneets
  kernel} code in the next time-step. The $x<$35 portion of the injected
spectrum from the previous time-step is directly evolved with
\textbf{Kompaneets kernel} at that time-step, giving the total $x<$35
spectrum at the end of the time-step as the sum of the outputs of two codes.
At the beginning of each time-step for processing by the
\textbf{Recoil-only} code, we set the $x<$35 portion of the spectrum to
zero so that there is no double counting and the recoil effect is not taken
into account twice. The \textbf{Kompaneets kernel} code therefore has full
control of $x<$35 spectrum evolution. The  $x>$35 portion of
\textbf{Recoil-only} code is evolved with full electromagnetic cascade
taken into account. For photons at $x$=35, \textbf{Recoil-only} prescription
in the Compton scattering is an excellent approximation. We have varied the
boundary point to $x$=30 and 25 and checked that it does not affect the
solution for $x\lesssim$20. To summarize, we evolve the spectrum of
$x\geq$35 with \textbf{Recoil-only} approximation for Compton scattering
and $x<$35 part of the spectrum is evolved with the full \textbf{Kompaneets
  kernel}. The flowchart illustrating the interface between the
two codes is shown in Fig. \ref{fig:flowchart}.  For the ease of numerical calculation, for evolution with the Kompaneets kernel, we have taken
the background electrons to be at the temperature of the undistorted CMB
Planck spectrum and all results shown are in this approximation. The electrons are actually at a slightly higher
temperature defined by the full CMB spectrum, including the distortions
(Eq.~\ref{eltemp}). Assuming the electrons to be at a lower temperature
will therefore add a small amount of cooling to the calculation equivalent to subtracting a
small thermal ($i,y,$ or $\mu$-type) spectral distortion. We can correct
for this at the end of the calculation by adding back the thermal
distortions calculated using energy conservation at each time-step to the total (thermal $+$
non-thermal) spectral distortion. Note that this does not
affect the pure non-thermal part of the spectral distortions (see
Sec. \ref{sec:dmdecay}).    \par
\hspace{1cm}
We consider 500 bins in x variable, with \textbf{Recoil-only} code. For one-time injection scenario, we start the code at redshift of energy injection and end at redshift z=1200 with time-steps of $\Delta z=100$. We have checked numerical convergence by varying the step sizes. For the \textbf{Kompaneets kernel} code, we have 1000 energy bins between $x$ of 0.01 to 35 and we track each collision of these low energy photons with the background electrons as described in Sec.~\ref{subsec:lowenergyph}. After recombination, efficiency of Comptonization decreases significantly. So, we go from redshift 1200 to present day in one step. There is no feedback from the \textbf{Kompaneets kernel} code to the \textbf{Recoil-only} code. This is an excellent approximation. We will compare the results obtained from the full code labelled \textbf{Kompaneets kernel} with those obtained by using just the \textbf{Recoil-only} code. \par
\hspace{1cm}
We consider monochromatic electron-only, electron-positron pair and photon-only injection. For electron energy injection, only its kinetic energy is dissipated as heat to the background electrons or as spectral distortion of the CMB photons. The energy dissipation of a positron is well approximated by the dissipation  of
same kinetic energy electron plus two photons at 511 keV. Let $f_X$ be the total kinetic energy density of injected electron-positron pairs with identical kinetic energies for both particles, $E_{e^{-}}=E_{e^{+}}$. The number density of positrons injected is equal to $\frac{f_X}{2E_{e^{+}}}$. Total 511 keV photon energy density will then be given by 2$\times f_X\times\frac{511 \mathrm{keV}
}{2E_{e^{+}}}$. This would be a big factor for a low energy positron i.e. with $E_{e^{+}}$ 10 keV while for $E_{e^{+}}>>m_{\mathrm{e}}$, it does not make a big difference.    
  \par
 In Fig.~\ref{fig:10kev100gevrecoil-only}, we show a few results for spectral distortions for one-time energy injection, with \textbf{Recoil-only} code used for the full spectrum including $x<$35 part, varying the kinetic energy of injected electrons and photons and compare them with the standard y-distortion. The energy injected is $10^{-5}$ times the energy density of the CMB at the injection redshift, $\rho_{\mathrm{CMB}}$. On the y-axis we plot the change in the intensity of the CMB from a Planck spectrum while on the x-axis the variable ($x=E_{\gamma}/kT$) is plotted. For all the initial energies, $y$-distortion created due to the heating of background baryon-photon plasma by low energy electrons and photons is of the order of $10^{-6}$ . We are interested in the comparison of the shape only. Therefore, we have only shown a scaled $y$-distortion spectrum for comparison. The generic feature of the plots is that the peak and the zero-point crossing are functions of the energy of the injected particle as well as whether the particle is an electron or a photon. There is also a long high energy tail with higher amplitude compared to the $y$-distortion case. \par
 \hspace{1cm}
 
 In Figs.~\ref{fig:10kev100gevkom},~\ref{fig:100mev1gevkomdelta1} and ~\ref{fig:1gev100gevkomdelta1}, we show the calculation with the \textbf{Kompaneets kernel} code used for the low energy ($x<$35) part of spectrum. As expected, the peak of the curve shifts to the right while the amplitude goes down as the total energy injected is held constant and the kinetic energy of the injected electrons is increased. For energies between 1 MeV to 10 MeV, most of the photons are at high $x$. This can be seen in Fig.~\ref{fig:1mev10mevrecoildelta} and \ref{fig:4mev10mevrecoildelta}. At about 8 MeV, One can see a turnaround of the photon spectrum. The IC scattered photons from the injected electrons give sufficient kinetic energy to the background electrons such that these secondary electrons themselves produce boosted but low energy CMB photons through IC scattering. So, there is again a rise in the amplitude of the spectrum at small $x$. From 20 MeV to 100 MeV, the peak of the distortions again starts to shift from left to right as the energy is increased and the whole cycle repeats. On close observation, it can be noticed that there is more than one such cycle. These are essentially the striping patterns seen in the plots of energy deposition curves in \citep{Slatyer:2015kla}. At about 10 GeV, the curves converge to a universal spectrum as was explained in Sec. \ref{subsec:enlossph}. \par
\hspace{1cm}
In Figs.~\ref{fig:10kev10mevepkomdelta1}, \ref{fig:10mev100gevepkomdelta1}, \ref{fig:10mevepkomdelta}, we compare the spectral distortion solution for the monochromatic electrons and photons.  For 10 keV photons, there is negligible $ntr$-type distortion  though there will be a $y$-type distortion.  For energies less than 10 GeV, distorted spectra of electron only and photon only injections are significantly different. At about 10 GeV, the spectra for the monochromatic electrons and photons become indistinguishable and the spectrum converges to a universal solution. This feature also holds true for energy deposition efficiency calculation around the recombination epoch \cite{Slatyer:2012yq}. In Fig. \ref{fig:e-e+vse-delta}, we have compared electron-positron pair injection with
electron only injection. For 10 keV, the spectrum is controlled by
511 keV photons while for 100 GeV the difference between the
electron-only and the electron-positron pair injection vanishes.
In Fig.~\ref{fig:10kev100kevvsz}, we compare the recoil only solution and full Kompaneets kernel solution for two low energy injected electrons of 10 keV and 100 keV by varying the redshift of energy injection while keeping the injected energy with respect to the CMB constant, $\frac{\rho_{inj}}{\rho_{
\mathrm{CMB}}}=10^{-5}$. In Fig.~\ref{fig:1gev100gevvsz}, we repeat the same for high energy electrons of 1 GeV and 100 GeV energies respectively. For all the cases, the \textbf{Recoil-only} solution gets sharper and sharper with increasing amplitude for increasing redshift. The photons at higher $x$ are being sent over to lower $x$ by recoil with rate proportional to square of $x$ (or frequency)  (see Eq. \ref{compton}). With higher redshift of energy injection, the photons move more and more to the left which explains the increasing sharpness. With the  \textbf{Kompaneets kernel}, the solution is less peaked as Doppler broadening smoothens out the spectrum. This broadening also moves the minimum point of the spectrum. The location of the minimum and the zero point crossing move in the same direction just like in Fig.~\ref{fig:ytomu1}. For the high energy case, there is a big increase in the amplitude of the spectrum going from redshift 10000 to 20000 both for \textbf{Recoil-only} and  \textbf{Kompaneets kernel} solution. This is because a significant proportion of photons and energy can be at quite large $x$ at redshift 10000 as these photons have not been able to move to low $x$. For redshift 20000, these photons suffer more scatterings due to high electron density and move to low $x$ and hence a higher amplitude of the spectral distortion. In Fig.~\ref{fig:10kev100gevvszkvsr}, we compare the \textbf{Recoil-only} and the \textbf{Kompaneets kernel} solution for different redshifts. For redshift 10000, there is not much difference between the \textbf{Recoil-only} and the \textbf{Kompaneets kernel} solution and the two curves overlap. As we go to higher redshifts, the electron density and hence the scattering rate of photons on electrons increases and Doppler broadening becomes important. The effect of Doppler broadening is clearly visible in Fig. \ref{fig:10kev100gevvszkvsr} and full \textbf{Kompaneets kernel} is needed for accurate calculation of the CMB spectral distortion.  \par
\hspace{1cm}

In Fig.~\ref{epheat}, we plot the fraction of the injected energy going to heat for energy injection at different redshifts for electron-only and photon-only energy injection. As the photons heat the background plasma through recoil, we have used the \textbf{Recoil-only} code for this calculation. Doppler broadening does not change the the amount of energy going to heat and stimulated scattering is not important at high $x$ where most of the heat transfer from photons to electrons takes place. Note that the \textbf{Recoil-only} code takes into account the average Doppler boost experienced by the photons and uses Eq. \ref{recoil}.  For higher and higher redshift injection, there is more and more energy going to heat as the scattering rate of the photons on the electrons is higher due to the  higher electron number density. For the 10 keV electron, a large fraction of energy goes to heat as its energy degrades to $\sim$keV range. For 10 keV photon also, almost all energy goes to heat. As the energy of the injected electron is increased, more and more low energy photons are produced which are inefficient in depositing their energy to the background plasma.  At about 10 MeV, the injected electron can boost a CMB photon to $\sim$10 keV which are very efficient in heating the medium. So, there is again a rise in heating efficiency. For even higher energetic injected electrons, the boosted CMB photon can impart sufficient energy to the background electron such that this background electron can again produce low energy photons through IC scattering. So, again the heating efficiency drops and the cycle is repeated again. As we go to higher and higher  energy, there will be a progressively broader photon spectrum as a result of the particle cascade. Some fraction of these photons deposit their energy efficiently  while rest of them can be inefficient. This results in progressive smoothening of the heating efficiency curve until it flattens out for energies $\gtrsim$GeV. High energy photons will produce high energy electrons and these electrons will produce the same pattern as described above. At very high energy, electron, positron, and photon injection are indistinguishable from each other. For higher and higher redshift, the CMB photons are at higher average  energy. So, an electron with lesser and lesser kinetic energy can boost a CMB photon to a particular energy. The collision rate of a photon with the background electrons and photons increases with increasing redshift due to the increase in their number density. This results in a progressive leftward shift of the heating efficiency curve. In Fig. \ref{fig:eheatx}, we have plotted the fractions of energy going to heat, photons with $x<=20$ and $x>$20 in the spectral distortions as a function of electron energy. The sum of these fractions adds upto one. Fraction of energy going to heat and photons with $x<=$20 are inversely related (i.e. minimum of heating efficiency is maximum of $x<=$20 photons) as these photons have higher chance of surviving as compared to high $x$ photons which deposit their energy as heat through recoil on the background electrons in Compton scattering. 

\subsection{Application to dark matter decay}\label{sec:dmdecay}
Before we conclude, let us apply the formalism developed in this paper to
the case of dark matter decay.  Any energy injection in the early
Universe will also affect the CMB anisotropy power spectrum. The CMB
anisotropies can get affected by the energy injection in two different
ways: (i) through direct ionization and excitation of recombining atoms by
the cascading high energy particles. (ii) the small modification of the
background CMB spectrum and change in the electron temperature to which the
recombination and photoionization rates are sensitive. For the dark matter
decaying during the recombination epoch, the mechanism (i) results in
change in the recombination history and provides stringent constraints
\cite{Poulin:2016anj}. For high redshift energy injection considered in
this paper, almost all of the energy is deposited into the CMB spectral
distortions much before recombination and no high energy particles are left
to affect recombination.   As far as mechanism (ii) is concerned, we already know from COBE that the spectral
distortions are very small and therefore the electron temperature cannot be
very different from the CMB temperature during recombination.  Therefore if
all the energy is injected before the recombination epoch, we expect the
CMB anisotropies to be almost unaffected and constraints from CMB
anisotropies to be extremely weak \cite{Poulin:2016anj}. For this reason,
CMB anisotropy constraints for decaying dark matter with lifetime less
than recombination epoch have not been considered by anyone so far. For
example, in \cite{Poulin:2016anj}, the authors have
calculated CMB anisotropy constraints for decaying dark matter lifetime 
$t_X\gtrsim 10^{13}~$s  (decay redshift $z_X\lesssim 1600$). We will
therefore only consider dark matter lifetimes much smaller
than the recombination time where the CMB ansiotropy limits are almost
non-existent and constraints from CMB spectral distortions become
important.  

The energy injection rate for particle decay  is given by
\citep{Chluba:2011hw,Chluba:2013wsa}, $\frac{dE}{dt}=f_X \Gamma
N_{\mathrm{H}} \exp(-\Gamma t)$ where $N_{\mathrm{H}}$ is the hydrogen
number density, the $f_X$ parameter contains the information about the dark
matter mass and its abundance with respect to the hydrogen, $\Gamma$ is the
inverse particle decay lifetime which can be converted  to a decay redshift  $z_X$. We assume that dark matter decays only to monochromatic electron-positron pairs or photons for simplicity without assuming any particular particle physics model of dark matter. We state our results in terms of the energy of decay products or injected particle, which in general may be different from the dark matter mass. In Fig. \ref{fig:decay}, we show the energy injection profile for dark matter decay and in Fig. \ref{fig:10kev100gevdm} the spectral distortion plots for dark matter decay with $z_X$=2.0$\times 10^{4}$ and $f_X$ such that if all injected energy ends up in heat then $y$-distortion will be $y$=1.5$\times10^{-5}$, the maximum allowed energy injection constrained by the COBE limits \cite{Fixsen1996}. This corresponds to decaying dark matter of fraction $10^{-4}$ compared to total dark matter provided it decays to electron-positron pairs or photons only. The lifetime corresponds to $t_X\sim 10^{11}$s.   As can be seen from the plots, the distortions are of the order of $\gtrsim10^{-6}$ and the shape is sensitive to the energy of the initial particle that is injected. The shape of the $ntr$-type distortion therefore contains information about the mass of the particle as well as the decay channels. These distortions if detected will therefore not only tell us that energy was injected into the CMB but can also tell us about the particle properties of the dark matter. We will consider in detail the applications to different specific dark matter models and the ability of the $ntr$-type distortions to distinguish between dark matter models in a future publication. We finally note that the amplitudes of the peaks of the $ntr$-type distortions are in general smaller than the case when we convert all the energy to $y$-type distortion. This is a generic feature of the $ntr$-type distortions. The more energetic electrons lose their energy to smaller number of photons but boost them to higher energies giving a high $x$ tail, with larger amplitude compared to the $y$-distortion.  
 \par
\begin{figure}[!p]
  \begin{subfigure}[b]{0.4\textwidth}
    \includegraphics[scale=0.8]{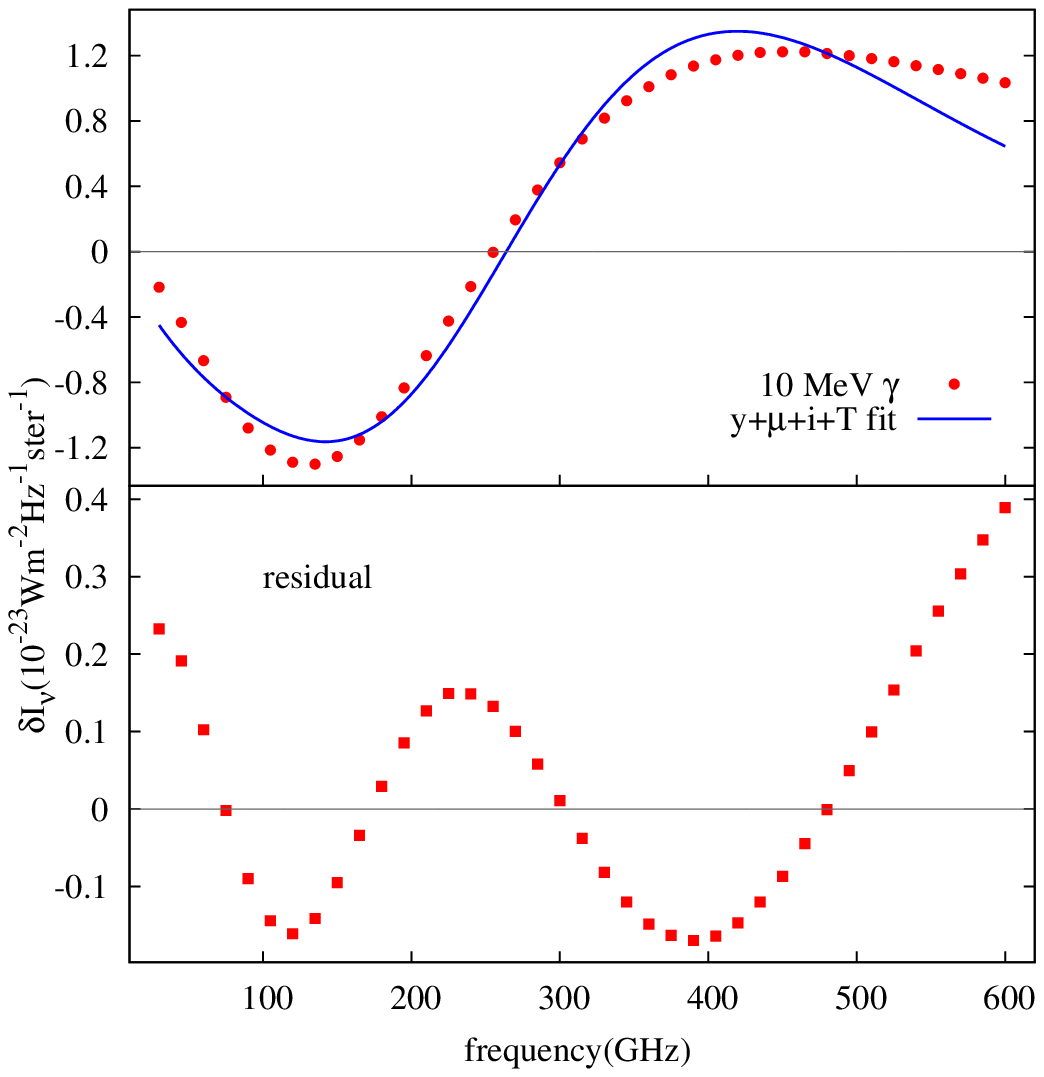}
     \label{fig:10mevgfit}
  \end{subfigure}\hspace{65 pt}
  \begin{subfigure}[b]{0.4\textwidth}
    \includegraphics[scale=0.8]{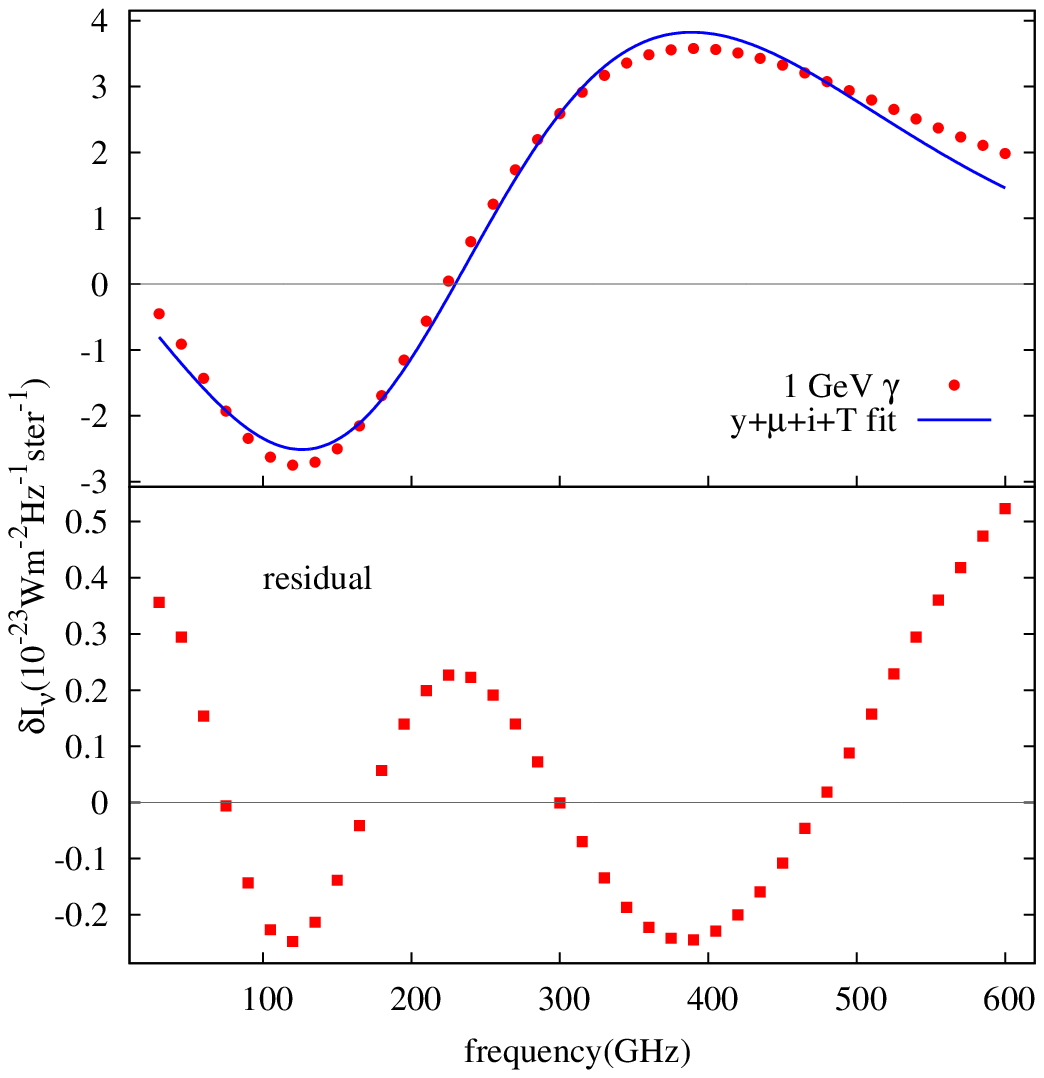}
    \label{fig:1gevgfit}
    \end{subfigure}\\
    
    \begin{subfigure}[b]{0.4\textwidth}
    \includegraphics[scale=0.8]{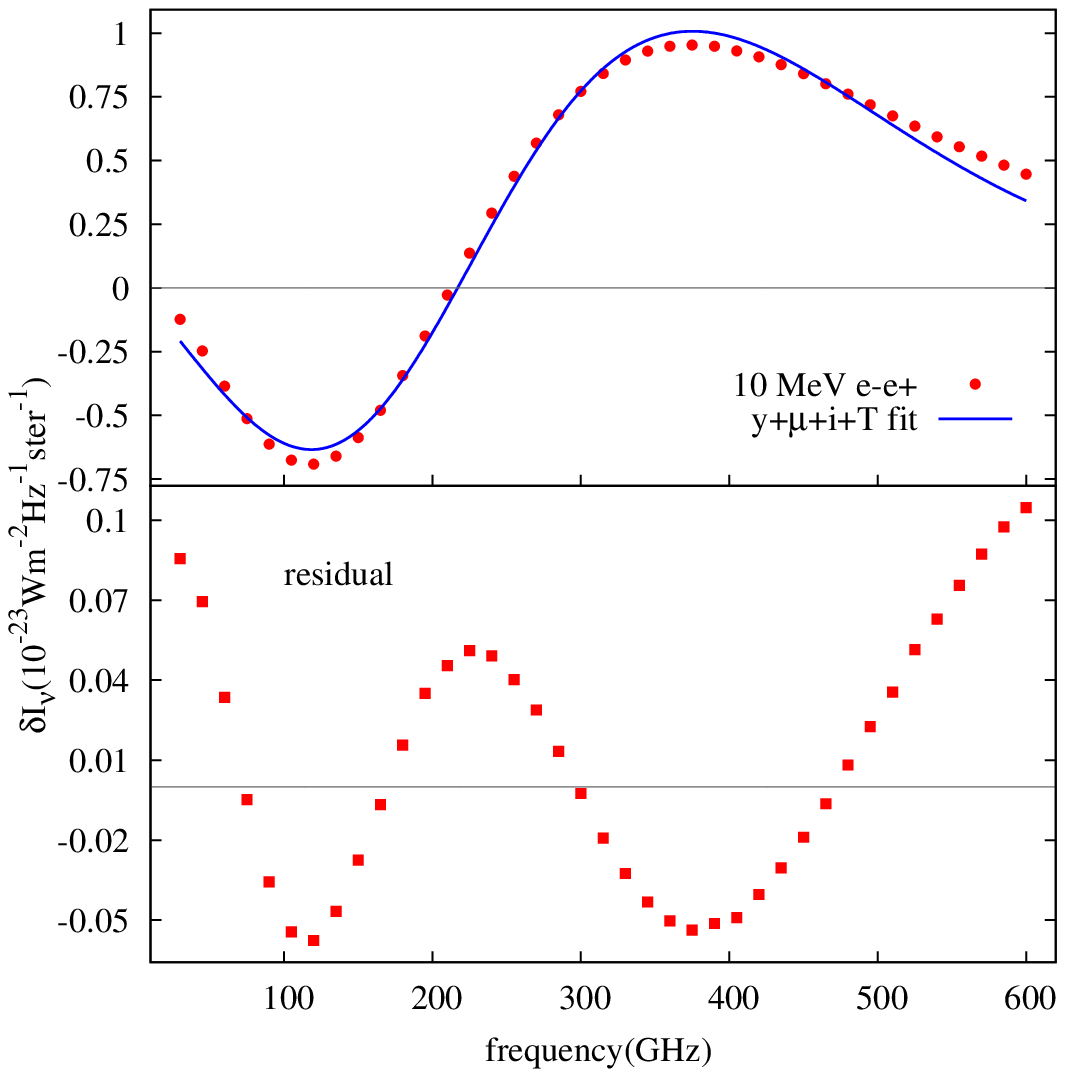}
  \label{fig:10mevefit}  
  \end{subfigure}\hspace{65 pt}
  \begin{subfigure}[b]{0.4\textwidth}
    \includegraphics[scale=0.8]{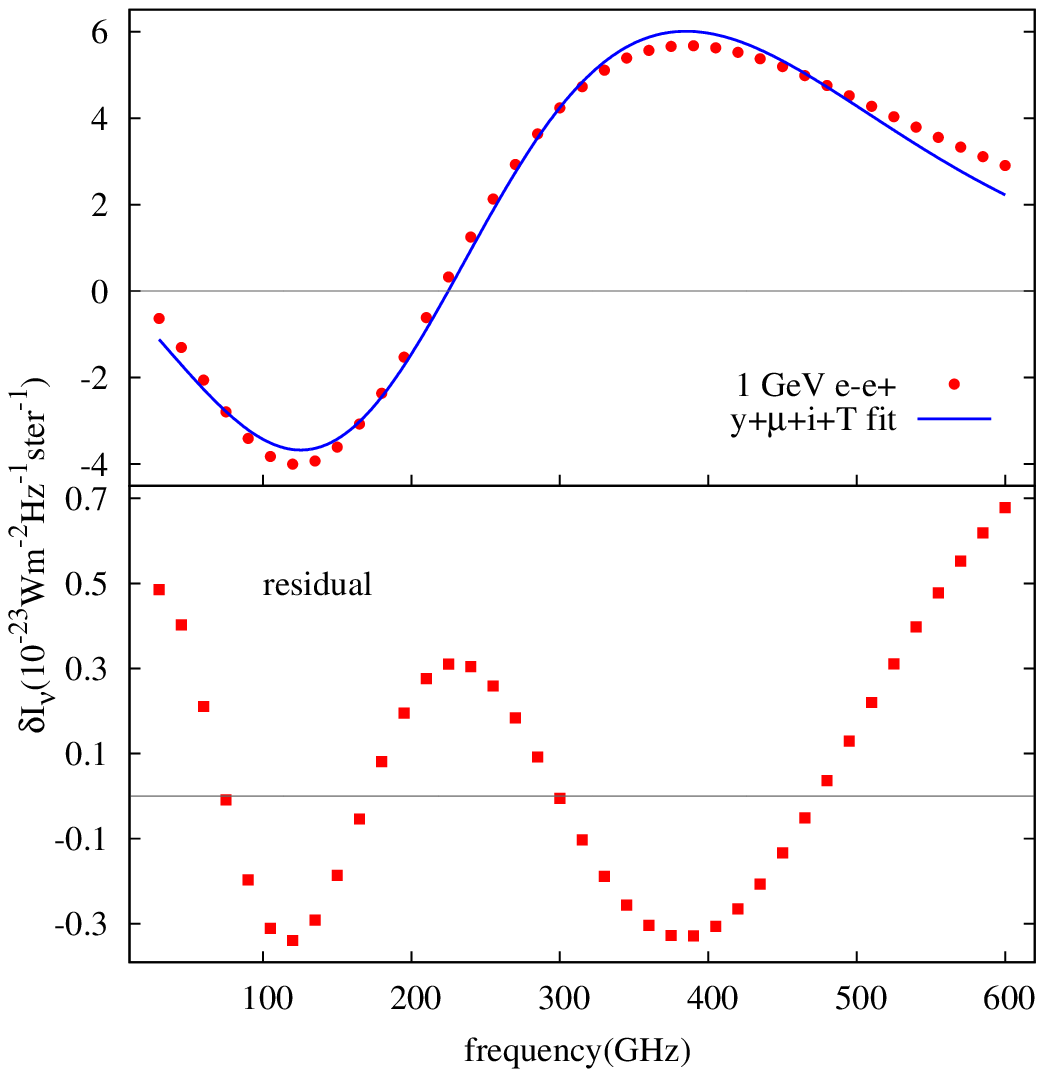}
   \label{fig:1gevefit}   
   \end{subfigure}
   \caption{Comparison of $ntr$-distortion for decaying dark matter 
     (decay redshift $z_X=2\times 10^4$) with best fit thermal distortion
     (y-type,i-type, $\mu$-type and temperature shift). The residuals can
     be considered to be 
     the pure non-thermal part of the spectral distortion.}
  \label{fig:fit}
\end{figure}

    Just as the $i$-type distortions are not orthogonal to $y$
   and $\mu$-type distortions \cite{Ks2013}, similarly the shape of the
   $ntr$-type distortion, although different,
   is not orthogonal to the thermal distortions. We can see how much
   different the $ntr$-type distortions are compared to the thermal
   distortions by trying to fit the thermal distortions to the $ntr$-type
   distortions for a PIXIE like experiment \cite{Pixie2011}. We fit the
   sum of $y$-type, $\mu$-type, $i$-type at redshift 20000, and a temperature
   shift (see Appendix \ref{app:spec})  with the amplitude of the four spectra as free parameters to the
   $ntr$-type distortion. For the fit we  sample the distortions at equally spaced PIXIE channels from 30
   GHz to 600 GHz with 15 GHz spacing between the channels. The results of
   such a fit are shown in Fig.\,\ref{fig:fit} for a few decay channels of
   decaying dark matter with decay redshift $z_X=2\times 10^4$. 
    We also show the residuals which are the  difference between the
    $ntr$-type distortion and the best fit thermal distortions. The
    residuals are typically $\gtrsim 10\%$ over most of the spectrum, and
    especially large at high frequencies. This is expected since the
    relativistic high
    energy cascading particles usually have a bigger high frequency tail in
    the spectral distortions compared to the non-relativistic thermal
    distortions. The presence of high frequency channels and efficient
    removal of high frequency foregrounds will therefore be important for
    detecting the $ntr$-type distortions in the future experiments.  

\clearpage
\section{\label{sec:disc}Discussion}
 We have calculated the spectral distortions of the CMB for energy injection at high redshifts in the form of high energy monochromatic electrons, positrons, or photons. We show that the usual assumption of \textbf{all of the injected energy dissipating as heat} in the background medium and producing a $y$-distortion is not correct. The spectral distortions created are sensitive to the energy and nature of the injected particle (electron, positron, or photon). We call these distortions non-thermal relativistic or $ntr$-type distortions to distinguish them from the other distortions that have been studied at these redshifts ($y$, $i$ and $\mu$-type) which are produced by interaction of the CMB with a non-relativistic and thermal distribution of electrons. The cascading high energy particles interact with the CMB, losing energy and creating distortions before the cascade reaches electrons with low enough energy that they thermalize and lose all of their energy as heat. So, the CMB spectral distortions retain some memory of the initial energy injection mechanism. This argument holds true for sub keV photons in the redshift range we are dealing with since they lose only a small fraction of their energy and can survive as a spectral distortion until today. The high energy photon collision processes produce large amount of energetic electrons by boosting the background electrons which by inverse Compton scattering boost the CMB photons to sub-keV energies. We have shown that the spectral distortion shape varies with the nature of the injected particle, and its energy, in the energy range keV $\lesssim E_{\mathrm{inj}}\lesssim$ 10 GeV and that the spectral distortion converges to a universal fixed spectrum at energies $\gtrsim$10 GeV becoming insensitive to the energy or the nature of the injected particles. \par
\hspace{1cm}

To evolve the full photon spectrum, we have developed a code that keeps track of photons taking into account all relevant collision processes in an expanding universe very much in the spirit of \citep{Slatyer:2009yq,Slatyer:2015kla,Kanzaki:2009hf}. However, in previous calculations, the background electrons were assumed to be at zero temperature. This is not a good assumption for photons of $x\sim$20 as the electrons have average energy $x\sim$1.5 and a non-negligible Maxwellian tail. For, a more accurate evolution of such low energy photons we have developed a new low energy code which takes  Doppler broadening, Doppler boosting and stimulated scattering into account. We have tested our low energy code by reproducing the calculation of \cite{Ks2013} by injecting a $y$-distortion at a particular redshift and evolving it with time to reach the equilibrium $\mu$ distortion. We use the kinetic equation of photon evolution  Eq. \ref{kinetic} to evolve the spectral distortion, which is equivalent to solving the Kompaneets equation in the Fokker-Planck approximation \citep{Ss2000}. With our code, we can track individual collisions of photons with the background electrons.\par
\hspace{1cm}
We have studied the case for dark matter decay with decay redshift of $z_X=20000$. For energy injection of $\frac{\delta\rho}{\rho_{\mathrm{CMB}}}\approx$6.0$\times 10^{-5}$, we show that the $ntr$-type distortion can be of the order of $\gtrsim10^{-6}$. For s-wave annihilation , there are stringent constraints on the annihilation cross-section from CMB anisotropy power spectrum observations \cite{Galli:2013dna}. So, we expect the signal from such annihilation mechanism, allowed by current data, to be lower than decay. We will explore in detail the ability of the $ntr$-type spectral distortions to distinguish between different energy injection scenarios and specific dark matter models in an upcoming paper. Our calculations in this paper already show that $ntr$-type distortions hold great promise. In effect, what we have calculated can be considered the \emph{Green's functions} or a complete set of solutions of the problem similar to the set of solutions for the $i$-type distortions calculated in \cite{Ks2013,Chluba:2013vsa}. The spectral distortion signal from any general injection scenario, be it a specific dark matter decay model or cosmic strings or any other new physics, can be calculated as a linear superposition of our solutions. We have shown that using just the total injected energy, which is the only parameter used for studying spectral distortions so far, is an oversimplification. The rich structure of the $ntr$-type spectral distortions contain much more information. The details of the energy injection mechanisms do matter and can be constrained by future observations of the CMB spectral distortions \cite{Pixie2011}.   
\section*{Acknowledgements}
We would like to thank Jens Chluba for comments on the manuscript. This work was supported by Science and Engineering Research Board (SERB) of Department of Science and Technology, Govt. of India  grant no. ECR/2015/000078. This work was also supported by MPG-DST partner group between Max Planck Institute for Astrophysics, Garching and Tata Institute of Fundamental Research, Mumbai funded by Max-Planck-Gesellschaft.    
\bibliographystyle{unsrtads}
\bibliography{specdist8}    
\appendix
\section{Collision cross-sections and energy loss rates for electrons and positrons}\label{app:elec}
\subsection{Coulomb scattering}
For energy loss of incident electrons with background electrons, we have used the formula \citep{Kanzaki:2008qb,Sng1971},
\begin{equation}
\frac{dE}{dt}=\frac{2.0\times 10^{-4}{n_{\mathrm{e}}}^{0.97}}{E^{0.44}}\left(\frac{E-E_e}{E-0.53E_e}\right)^{2.36}\mathrm{eVs^{-1}},
\end{equation}
where $E$ is the incident electron kinetic energy in eV, $E_{\mathrm{e}}$ is the background electron energy in eV, and $n_{\mathrm{e}}$ is the background electron number density in $\mathrm{cm^{-3}}$.
\subsection{Annihilation of positron with electron}
The electron-positron annihilation cross-section in the non-relativistic regime taking into account Coulomb correction is given by \cite{Ll1958,Cjrw1976},
\begin{equation}
\mathrm{\sigma}=\pi {r_o}^2\frac{2\pi\alpha(\frac{1}{\beta})^2}{1-\mathrm{\exp}(-2\pi\alpha\frac{1}{\beta})}.
\end{equation}
Under the assumption that initial particles are plane waves, the annihilation cross-section is given by \cite{H1954,Cjrw1976},
\begin{equation}
\mathrm{\sigma}=\frac{\pi {r_o}^2}{\gamma+1}\left[\frac{\gamma^2+4\gamma+1}{\gamma^2-1}\mathrm{ln}(\gamma+\sqrt(\gamma^2-1))-\frac{\gamma+3}{\sqrt(\gamma^2-1)}\right],
\end{equation}
where $\beta$ and $\gamma$ are the boost factor and Lorentz factor of positron and $r_o$ is the classical electron radius. 
\subsection{Inverse Compton scattering of a relativistic electron}
The energy-loss rate for inverse Compton scattering of an electron with the CMB in Thomson approximation is given by \citep{Gb1970,Kanzaki:2008qb},
\begin{equation}
\frac{dE}{dt}=\frac{4}{3}\sigma_{\mathrm{T}}cU_{\mathrm{CMB}}\gamma^2_e\beta^2_e,
\label{IC}
\end{equation}
where $U_{\mathrm{CMB}}$ is the energy density of CMB.
For the spectrum of inverse Compton scattered photon from a monoenergetic relativistic electron, we use the formula for the frequency distribution function \cite{W1979,R1995}. The analytic formula is given by, \citep{Ek2000},
\begin{multline}
P(t;p)=\frac{-3\left|(1-t)\right|}{32p^6t}[1+(10+8p^2+4p^4)t+t^2]\\+\frac{3(1+t)}{8p^5}\left[\frac{3+3p^2+p^4}{\sqrt{1+p^2}}-\frac{3+2p^2}{2p}(2\arcsinh{p}-\left| \ln{t}\right|)\right],
\end{multline}
where $t=\frac{\nu'}{\nu}$, $\nu$ is the frequency of unscattered photon, $\nu'$ is the frequency of scattered photon, $p=\beta_e\gamma_e$ and $\gamma_e$ is the Lorentz factor of electron. The maximal frequency shift is given by, $  \left| \ln{t}\right|< 2\arcsinh{p}$
\subsection{Inverse Compton scattering of an electron in ultra-relativistic limit}
In the redshift range we are interested in, the electron rest frame energy of a  CMB photon for an ultra-relativistic electron (of energy $\sim$ GeV or more) is comparable or more than the electron mass. So, in this case Klein-Nishina cross-section has to be used. We use the formula for double differential spectrum  \cite{Gb1970},
\begin{multline}
\frac{d^2N}{dtdE_1}=\frac{2\pi {r_0}^2m_{\mathrm{e}}c^3}{\gamma_e}\frac{n(\epsilon)d\epsilon}{\epsilon}\left[2q\ln{q}+(1+2q)(1-q)+0.5\frac{(\Gamma q)^2}{1+\Gamma q}(1-q)\right],
\end{multline}
where $\Gamma=\frac{4\epsilon\gamma_e}{m_{\mathrm{e}}c^2}$, $q=\frac{E_1}{\Gamma(1-E_1)}$, $\epsilon$ is the unscattered photon energy, $n(\epsilon)$ is the incident photon distribution, $E_1=\frac{\epsilon_1}{\gamma_em_{\mathrm{e}}c^2}$ and $\epsilon_1$ is the energy of scattered photon. 
\section{Collision cross-section and energy loss rate for photon}
\label{app:photon}
\subsection{Photo-ionization}
The photo-ionization cross-section is given by \citep{Slatyer:2009yq,Sz1989},
\begin{equation}
\sigma=\frac{2^9\pi^2 {r_0}^2}{3\alpha^3}{\left(\frac{E_{\mathrm{thres}}}{E}\right)}^4\frac{\mathrm{exp}(-4\eta \arctan{(1/\eta)})}{1-\exp{(-2\pi\eta)}},
\end{equation}
where $\eta=\frac{1}{\left((\frac{E}{E_{\mathrm{thres}}})-1\right)^{1/2}}$. $r_0$ is the electron radius. $E_{\mathrm{thres}}$=13.6eV for hydrogen and $\alpha$ is the fine structure constant.
\subsection{Compton scattering}
For non-relativistic Compton scattering, the energy-loss rate is given by \citep{Sz1990},
\begin{equation}
\frac{dE}{dt}=n_{\mathrm{e}}\sigma_{\mathrm{T}}c\frac{(h\nu)^2}{m_{\mathrm{e}}}.
\label{compton}
\end{equation}
The differential Klein-Nishina cross-section is given by \citep{Kn1928,Kanzaki:2008qb,H1954},
\begin{equation}
\frac{d\sigma}{d\epsilon}=\frac{3\sigma_{\mathrm{T}}}{8}\frac{m_{\mathrm{e}}}{(h\nu)^2}\left[\frac{h\nu}{\epsilon}+\frac{\epsilon}{h\nu}+(\frac{m_{\mathrm{e}}}{\epsilon}-\frac{m_{\mathrm{e}}}{h\nu})^2-2(\frac{m_{\mathrm{e}}}{\epsilon}-\frac{m_{\mathrm{e}}}{h\nu})\right],
\end{equation}
for $\frac{m_{\mathrm{e}}}{m_{\mathrm{e}}+2h\nu}h\nu\leq\epsilon\leq h\nu$, where $h\nu$ is the incident photon energy, $\epsilon$ is the scattered photon energy, $\sigma_{\mathrm{T}}$ is the Thomson cross-section and $m_{\mathrm{e}}$ is the electron mass.
\subsection{Pair production on matter}
For pair production on electron and ionized nuclei, we use the formula \citep{Slatyer:2009yq,Mok1969},
\begin{equation}
\sigma=\alpha {r_0}^2Z^2\left(\frac{28}{9}\ln(2E_{\gamma})-\frac{218}{27}\right),
\end{equation}
where $\alpha$ is the fine structure constant, $r_0$ is the electron radius and $E_{\gamma}$ is the photon energy in units of electron mass.
The spectrum of positrons is given by,
\begin{equation}
\frac{d\sigma}{dE_+}=4\alpha Z^2{r_0}^2\frac{{E_+}^2+{E_-}^2+\frac{2}{3}E_{+}E_{-}}{{E_{\gamma}}^3}\left[\mathrm{ln}(\frac{2E_{+}E_{-}}{E_{\gamma}})-\frac{1}{2}\right],
\end{equation}
where $E_{+}, E_{-}$, and $E_{\gamma}$ are the positron, electron and initial photon energies respectively in units of electron mass. 
\subsection{Photon-photon elastic scattering on CMB}
The photon-photon scattering rate for an incident photon of energy $E_{\gamma}$ on CMB is given by \citep{Kanzaki:2008qb,Sz1990},
\begin{equation}
p(E_\gamma)=3.33\times 10^{11}\left(\frac{T_{\mathrm{CMB}}}{m_{\mathrm{e}}}\right)^6\left(\frac{E_{\gamma}}{m_{\mathrm{e}}}\right)^3 s^{-1}.
\end{equation}
Normalized distribution of secondary photons (boosted CMB photon and degraded original photon) is given by,
\begin{equation}
p({E_{\gamma}}',E_{\gamma})=\frac{20}{7}\frac{1}{E_{\gamma}}\left[1-\frac{E'_{\gamma}}{E_{\gamma}}+(\frac{E'_{\gamma}}{E_{\gamma}})^2\right]^2,
\end{equation}
where $0\leq E'{_\gamma}\leq E_{\gamma}$ is the energy of secondary photon.
\section{Thermal spectral distortions}\label{app:spec}
The intensity for $y$-distortion is given by \cite{Sz1969},
\begin{equation}
I_{y}(\nu)=A_{y}\frac{2h\nu^3}{c^2}\frac{xe^x}{(e^x-1)^2}\left[x\frac{e^x+1}{e^x-1}-4.0\right].
\end{equation}
The intensity for $\mu$-distortion is given by \cite{Sz19701,Is19751},
\begin{equation}
I_{y}(\nu)=A_{\mu}\frac{2h\nu^3}{c^2}\frac{e^x}{(e^x-1)^2}\left[\frac{x}{2.19}-1.0\right].
\end{equation}
The intensity for $i$-distortion is given by,
\begin{equation}
I_{i}(\nu)=A_{i}F_{\nu}(z)
\end{equation}
which has to be calculated numerically as was shown in Fig.\,\ref{fig:ytomu1}.
The intensity for temperature shift is given by,
\begin{equation}
I_{T}(\nu)=\Delta T\frac{\partial I_{\rm pl}}{\partial T}=\Delta
T\frac{2h\nu^3}{c^2}\frac{xe^x}{(e^x-1)^2},
\end{equation}
where $I_{\rm pl}$ is the Planck spectrum.
We approximate the $ntr$-type distortion by a linear combination of these
spectra and fit for the amplitudes $A_y, A_{\mu}, A_i$, and $\Delta T$ to $ntr$-type
distortions in Sec. \ref{sec:dmdecay}.
\end{document}